\documentclass[journal]{vgtc}         %
\ifpdf%
  \pdfoutput=1\relax                   %
  \usepackage{graphicx}                %
  \DeclareGraphicsExtensions{.pdf,.png,.jpg,.jpeg} %
\else%
  \usepackage{graphicx}                %
  \DeclareGraphicsExtensions{.eps}     %
\fi%

\graphicspath{{figures/}{pictures/}{images/}{./}} %

\usepackage{microtype}                 %
\PassOptionsToPackage{warn}{textcomp}  %
\usepackage{textcomp}                  %
\usepackage{mathptmx}                  %
\usepackage{times}                     %
\usepackage{cite}                      %
\usepackage{tabu}                      %
\usepackage{booktabs}                  %

\usepackage{hhline}
\usepackage{multirow}
\usepackage{colortbl}
\usepackage[table]{xcolor}

\usepackage{enumitem}
\usepackage{tikz-cd}
\usepackage{wrapfig}
\usepackage{adjustbox}

\usepackage{tikz}
\usetikzlibrary{arrows,arrows.meta,patterns,calc,shapes,decorations.pathreplacing}
\usepackage{etoolbox}
\usepackage{ifthen}
\makeatletter
\patchcmd\@hex@@Hex{f\else}{F\else}{}{}
\makeatother
\usepackage{rotating}
\newcommand{\boundellipse}[3]
{(#1) ellipse (#2 and #3)
}

\definecolor{branchColor1}{HTML}{1F77B4} 
\definecolor{branchColor2}{HTML}{AEC7E8} 
\definecolor{branchColor3}{HTML}{2CA02C} 
\definecolor{branchColor4}{HTML}{98DF8A} 
\definecolor{branchColor7}{HTML}{D62728} 
\definecolor{branchColor5}{HTML}{FFBB78} 
\definecolor{branchColor6}{HTML}{FF7F0E} 
\definecolor{branchColor8}{HTML}{9C00FF} 
\definecolor{branchColor9}{HTML}{D189FF} 

\definecolor{grayC}{HTML}{CCCCCC}
\definecolor{grayB}{HTML}{BBBBBB}
\definecolor{grayA}{HTML}{AAAAAA}
\definecolor{gray9}{HTML}{999999}
\definecolor{gray8}{HTML}{888888}
\definecolor{gray7}{HTML}{777777}
\definecolor{gray6}{HTML}{666666}
\definecolor{gray5}{HTML}{555555}
\definecolor{gray4}{HTML}{444444}

\definecolor{filtered}{HTML}{BBBBBB}
\definecolor{filtered2}{HTML}{CCCCCC}
\definecolor{filtered3}{HTML}{F5F5F5}

\definecolor{lm2}{HTML}{6FB0E7}
\definecolor{lm1}{HTML}{2484D6}
\definecolor{l1}{HTML}{BD0026}
\definecolor{l2}{HTML}{F03B20}
\definecolor{l3}{HTML}{FD8D3C}
\definecolor{l4}{HTML}{FED976}
\definecolor{l5}{HTML}{FFFFB2}

\definecolor{colorBrewerC12_0}{HTML}{a6cee3}
\definecolor{colorBrewerC12_1}{HTML}{1f78b4}
\definecolor{colorBrewerC12_2}{HTML}{b2df8a}
\definecolor{colorBrewerC12_3}{HTML}{33a02c}
\definecolor{colorBrewerC12_4}{HTML}{fb9a99}
\definecolor{colorBrewerC12_5}{HTML}{e31a1c}
\definecolor{colorBrewerC12_55}{HTML}{a40013}
\definecolor{colorBrewerC12_6}{HTML}{fdbf6f}
\definecolor{colorBrewerC12_7}{HTML}{ff7f00}
\definecolor{colorBrewerC12_8}{HTML}{cab2d6}
\definecolor{colorBrewerC12_9}{HTML}{6a3d9a}
\definecolor{colorBrewerC12_10}{HTML}{ffff99}
\definecolor{colorBrewerC12_11}{HTML}{b15928}
\definecolor{colorBrewerC12_12}{HTML}{595959}
\definecolor{colorBrewerC12_13}{HTML}{e80500}

\definecolor{ng3dt1}{HTML}{3182bd}
\definecolor{ng3dt2}{HTML}{6baed6}
\definecolor{ng3dt3}{HTML}{9ecae1}

\definecolor{ng3dn1}{HTML}{cc0000}
\definecolor{ng3dn2}{HTML}{d96666}
\definecolor{ng3dn3}{HTML}{ffabab}

\tikzstyle{contourNumber0} = [fill=white, draw=black, circle, inner sep=0pt]
\tikzstyle{contourNumber1} = [fill=white, draw=black, circle, inner sep=0.5pt]
\tikzstyle{contourNumber2} = [fill=white, draw=black, circle, inner sep=0.04em]

\tikzstyle{veryThinEdge} = [line width=0.01cm]
\tikzstyle{thinEdge} = [line width=0.03cm]
\tikzstyle{mediumEdge} = [line width=0.04cm]
\tikzstyle{thickEdge} = [line width=0.1cm]
\tikzstyle{dashedEdge} = [dashed, gray]

\tikzstyle{thickNode} = [shape=circle,draw=black,fill=black, inner sep=0pt, minimum size=4]
\tikzstyle{filteredNode} = [shape=circle,draw=filtered,fill=filtered, inner sep=0pt, minimum size=4]

\tikzstyle{arrow} = [-{Stealth[scale=1.5]}]
\tikzstyle{arrow2} = [-{Stealth[scale=1.1]}]

\definecolor{rLevel0}{HTML}{a93838}
\definecolor{rLevel1}{HTML}{e3aeae}
\definecolor{bLevel0}{HTML}{3e7fa4}
\definecolor{bLevel1}{HTML}{b1cfe1}
\definecolor{gLevel0}{HTML}{3ea440}
\definecolor{gLevel1}{HTML}{b1e1b2}

\definecolor{colorTheme_00}{HTML}{a71d44}
\definecolor{colorTheme_01}{HTML}{dd6767}

\definecolor{colorTheme_10}{HTML}{777777}
\definecolor{colorTheme_11}{HTML}{bbbbbb}

\definecolor{colorTheme_20}{HTML}{f27d00}
\definecolor{colorTheme_21}{HTML}{f1b473}
\definecolor{colorTheme_22}{HTML}{edc293}
\definecolor{colorTheme_23}{HTML}{f5cca1}

\definecolor{colorTheme_30}{HTML}{3ea440}

\definecolor{colorTheme_40}{HTML}{1a769c}
\definecolor{colorTheme_41}{HTML}{83a6c1}

\definecolor{simplexBG}{HTML}{ececec}

\definecolor{highlight0_r}{HTML}{e3aeae}
\definecolor{highlight1_r}{HTML}{a93838}

\definecolor{highlight0_b}{HTML}{b1cfe1}
\definecolor{highlight1_b}{HTML}{3e7fa4}

\tikzstyle{simplexVertex}=[draw=black, thinEdge, fill=white, shape=circle, minimum size=7pt, inner sep=2pt]
\tikzstyle{highlightedSimplexVertexR}=[draw=colorTheme_00, line width=0.75mm, fill=colorTheme_01, shape=circle, minimum size=1.3em, inner sep=2pt]
\tikzstyle{highlightedSimplexVertexB}=[draw=colorTheme_20, line width=0.75mm, fill=colorTheme_23, shape=circle, minimum size=1.3em, inner sep=2pt]
\tikzstyle{highlightedSimplexVertexG}=[draw=colorTheme_10, line width=0.75mm, fill=colorTheme_11, shape=circle, minimum size=1.3em, inner sep=2pt]

\tikzstyle{simplexEdge}=[draw=black, line width=0.1mm]
\tikzstyle{highlightedSimplexEdgeR}=[draw=colorTheme_00, thickEdge]
\tikzstyle{highlightedSimplexEdgeB}=[draw=colorTheme_20, thickEdge]
\tikzstyle{highlightedSimplexEdgeG}=[draw=colorTheme_10, thickEdge]

\tikzstyle{simplexTriangle}=[fill=simplexBG, draw=black, line width=0.1mm]
\tikzstyle{highlightedSimplexTriangleR}=[fill=rLevel1, draw=black, line width=0.1mm]
\tikzstyle{highlightedSimplexTriangleB}=[fill=bLevel1, draw=black, line width=0.1mm]
\tikzstyle{highlightedSimplexTriangleG}=[fill=gLevel1, draw=black, line width=0.1mm]

\tikzstyle{persistenceCurveMarker} = [circle, draw=rLevel0, line width=2pt, inner sep=0.2em]
\tikzstyle{persistenceCurveArrow} = [draw=black,arrow,densely dashed, line width=0.6pt]

\usepackage{amssymb}
\usepackage{amsmath}
\usepackage{amsbsy}
\usepackage{amsthm}
\usepackage{dsfont}
\usepackage{thmtools}
\usepackage{graphics}

\DeclareMathAlphabet{\mathcal}{OMS}{cmsy}{m}{n}

\newcounter{definitions}
\newtheorem{definition}[definitions]{Definition}

\makeatletter
\def\ll@definition{%
  \protect\numberline{\csname the\thmt@envname\endcsname}%
  \ifx\@empty\thmt@shortoptarg
    \thmt@thmname
  \else
    \thmt@shortoptarg
  \fi
}
\makeatother

\usepackage[ruled,vlined,linesnumbered]{algorithm2e}
\usepackage{algpseudocode}
\newcommand{\algorithmHSpaceDefault}{0.4cm}
\newcommand{\algorithmHSpaceValue}{\algorithmHSpaceDefault}
\newcommand{\algorithmHSpace}{\hspace*{\algorithmHSpaceValue}}
\SetAlCapHSkip{0em}
\newlength{\maxwidth}
\DontPrintSemicolon
\newcommand{\algAM}[2]%
{\makebox[\maxwidth][l]{$#1$}\algorithmHSpace$\leftarrow\;#2$}
\newcommand{\algAT}[2]%
{\makebox[\maxwidth][l]{$#1$}\algorithmHSpace$\leftarrow\;$#2}
\SetArgSty{textnormal}

\newcommand{\domain}{\mathcal{M}}
\newcommand{\vertices}{\mathcal{V}}
\newcommand{\range}{\mathbb{R}}
\newcommand{\sublevelset}[1]{{#1}^{-1}_{-\infty}}
\newcommand{\superlevelset}[1]{{#1}^{-1}_{+\infty}}
\newcommand{\superlevelsetcomponent}[2]{\domain_{#1}^{#2}}
\newcommand{\Star}{St}
\newcommand{\Link}{Lk}
\newcommand{\simplex}{\sigma}
\newcommand{\face}{\tau}
\newcommand{\lowerlink}{\Link^{-}}
\newcommand{\upperlink}{\Link^{+}}
\newcommand{\Index}{\mathcal{I}}
\newcommand{\order}[1]{\hat #1}
\newcommand{\Natural}{\mathbb{N}}
\newcommand{\criticalSet}{\mathcal{C}}

\newcommand{\preserve}[2]{\overline{\criticalSet_{#1}^{#2}}}
\newcommand{\speedup}[1]{$\times$\hspace*{-0.1em}$#1$}

\onlineid{1057}

\vgtccategory{Research}
\vgtcpapertype{Algorithm/Technique}

\title{Localized Topological Simplification of Scalar Data}

\author{
Jonas Lukasczyk, Christoph Garth, Ross Maciejewski, and Julien Tierny\vspace*{0.8em}
}
\authorfooter{
\item
 Jonas Lukasczyk and Ross Maciejewski are with Arizona State University.\\E-mail: jl@jluk.de and rmacieje@asu.edu.
\item
 Christoph Garth is with Technische Universit\"at Kaiserslautern.\\E-mail: garth@cs.uni-kl.de.
\item
 Julien Tierny is with Sorbonne Universit\'e and CNRS.\\E-mail: julien.tierny@sorbonne-universite.fr.
}

\shortauthortitle{Lukasczyk et al. \MakeLowercase{\textit{et al.}}: Localized Topological Simplification}

\abstract{
    This paper describes a localized algorithm for the topological simplification of scalar data, an essential pre-processing step of topological data analysis (TDA).
    Given a scalar field~$f$ and a selection of extrema to preserve, the proposed localized topological simplification (LTS) derives a function~$g$ that is close to~$f$ and only exhibits the selected set of extrema.
    Specifically, sub- and superlevel set components associated with undesired extrema are first locally flattened and then correctly embedded into the global scalar field, such that these regions are guaranteed---from a combinatorial perspective---to no longer contain any undesired extrema.
    In contrast to previous global approaches, LTS only and independently processes regions of the domain that actually need to be simplified, which already results in a noticeable speedup.
    Moreover, due to the localized nature of the algorithm, LTS can utilize shared-memory parallelism to simplify regions simultaneously with a high parallel efficiency ($70\%$).
    Hence, LTS significantly improves interactivity for the exploration of simplification parameters and their effect on subsequent topological analysis.
    For such exploration tasks, LTS brings the overall execution time of a plethora of TDA pipelines from minutes down to seconds, with an average observed speedup over state-of-the-art techniques of up to \speedup{36}.
    Furthermore, in the special case where preserved extrema are selected based on topological persistence, an adapted version of LTS partially computes the persistence diagram and simultaneously simplifies features below a predefined persistence threshold.
    The effectiveness of LTS, its parallel efficiency, and its resulting benefits for TDA are demonstrated on several simulated and acquired datasets from different application domains, including physics, chemistry, and biomedical imaging.
}


\CCScatlist{ %
}

\teaser{
    \hspace*{-0.5em}
    \begin{tikzpicture}

        \node at (0,0)
{\includegraphics[width=0.5\textwidth]{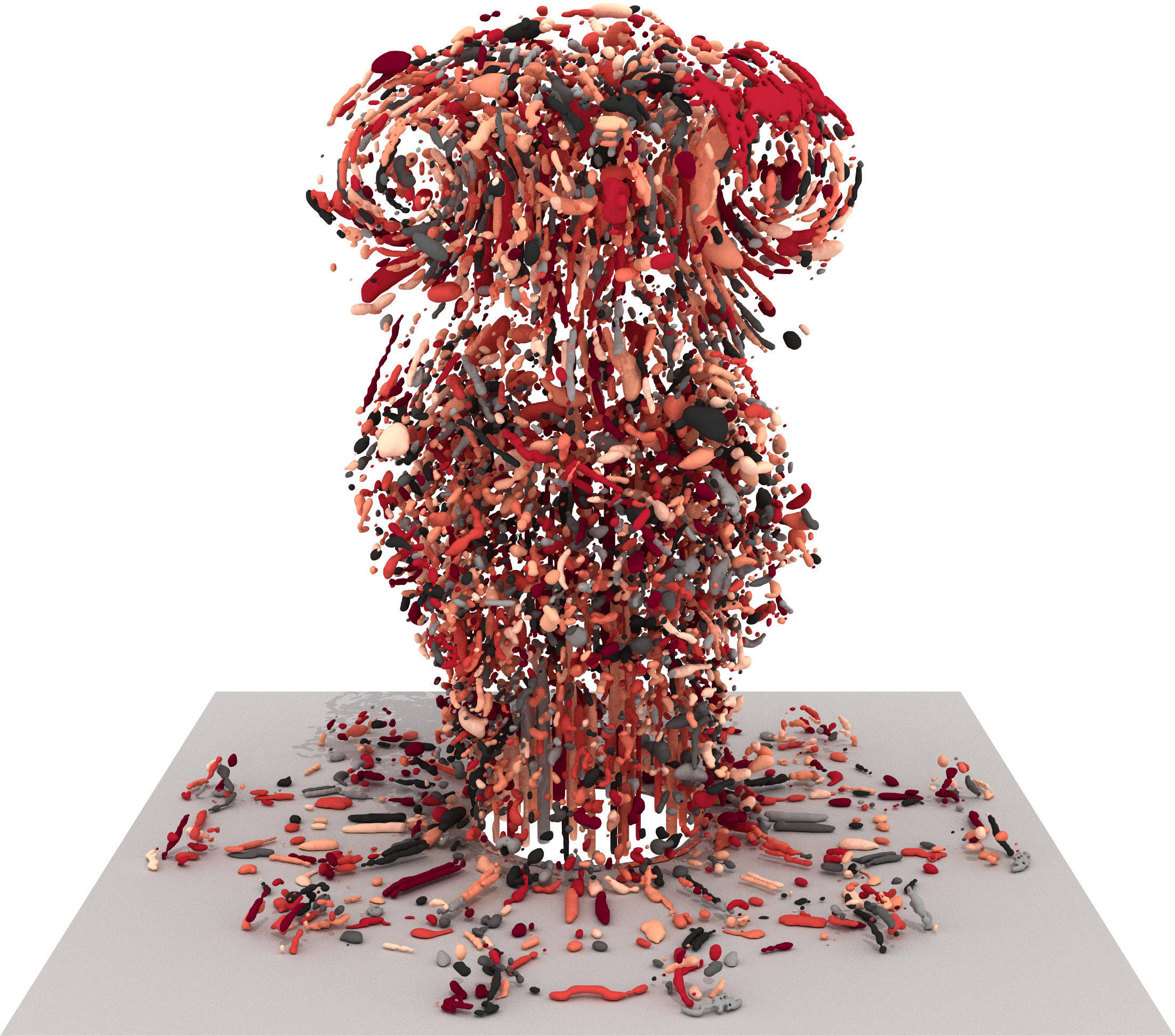}};
        \node at (8.2,0)
{\includegraphics[width=0.5\textwidth]{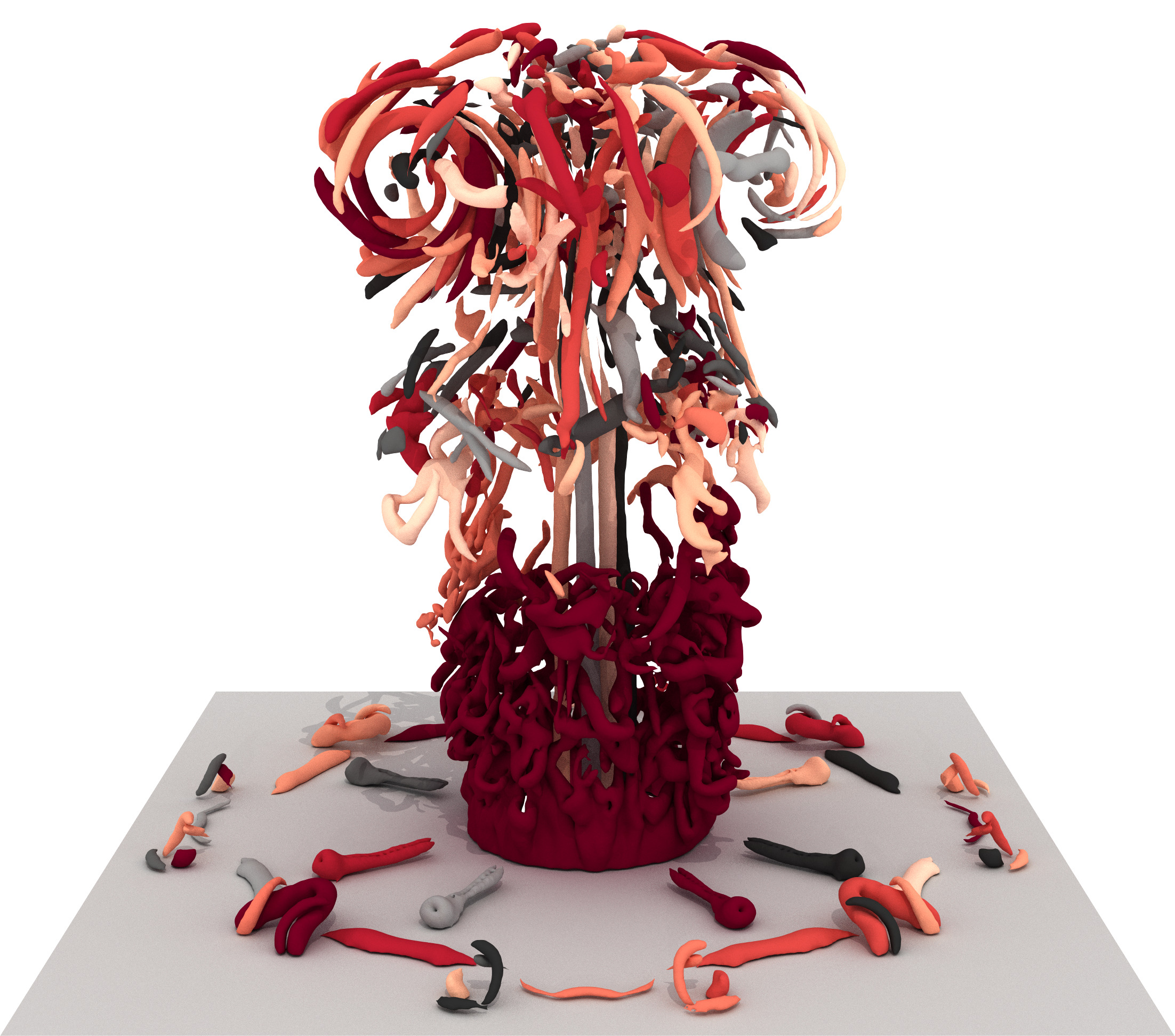}};

        \newcommand{\teaserLineCenter}{0.3}

        \node at (4.1,\teaserLineCenter+0.75) {\parbox{16em}{\centering \large
removing $\mathbf{20k}$ of $\mathbf{21k}$ maxima\\in a $\mathbf{512^3}$ scalar
field\\takes $\mathbf{7s}$ with \textbf{LTS}}};
        \draw[arrow,thick] (1.9,\teaserLineCenter) -- (6.3,\teaserLineCenter);
        \node at (4.1,\teaserLineCenter-0.6) {\parbox{12em}{\centering \large
        and $\mathbf{165s}$ with a\\state-of-the-art approach}};
    \end{tikzpicture}

    \mycaption{
        Illustration of the impact of localized topological simplification (LTS) on the resulting topology-based feature characterization of individual vortices (colored regions) in a computational fluid dynamics simulation~\cite{jetData}.
        The simulation models the injection of a jet into a medium at rest ($512^3$ voxels) where individual vortices are characterized as regions around vorticity magnitude maxima.
        By selecting maxima that need to be removed via different persistence thresholds for different parts of the domain, it is possible to suppress noise, and to combine multiple small vortices into larger regions\jonasRevision{; allowing users to explore different merging thresholds}.
        In this example, even for the aggressive removal of $20k$ out of $21k$ maxima, LTS only needs to process $1\%$ of the scalar field.
        Moreover, due to the algorithm's localized nature, LTS can utilize shared-memory parallelism to outperform previous approaches by an order of magnitude.
        In this example, exploring a different selection of maxima with the simplification algorithm available in the Topology ToolKit~\cite{ttk} takes roughly \julien{$165s$}, whereas LTS only requires \julien{$7s$}.
        This significant speedup enables a higher level of interactivity for data visualization and analysis, which is essential if simplification parameters and their effect are not known a priori.%
        \vspace*{1em}
    }
    \label{fig_teaser}
}

\vgtcinsertpkg

\begin{document}

\newcommand{\mycaption}[1]{\vspace*{-1em}\caption{#1}\vspace{-0.25em}}

\renewcommand{\sectionautorefname}{Sec.}
\renewcommand{\subsectionautorefname}{Sec.}
\renewcommand{\subsubsectionautorefname}{Sec.}
\renewcommand{\figureautorefname}{Fig.}
\renewcommand{\equationautorefname}{Eq.}
\renewcommand{\tableautorefname}{Tab.}

\newcommand{\julien}[1]{\textcolor{black}{#1}}
\newcommand{\julienRevision}[1]{\textcolor{black}{#1}}
\newcommand{\jonas}[1]{\textcolor{black}{#1}}
\newcommand{\jonasRevision}[1]{\textcolor{black}{#1}}

\firstsection{}
\maketitle

\section{Introduction}
\label{sec_introduction}

\begin{figure*}
  \centering
  \input{figures/cells/cellsIntro.tex}

  \mycaption{
    Extraction of plant cell nuclei depicted in a microscopy
image~\cite{cellImage} (a) via sublevel set components of the color brightness
scalar field that has been simplified based on different strategies (b-d).
    Without any simplification (b), the extracted cells contain numerous false
positives in the form of undesired components that result from non-persistent
extrema.
    Bridging (c) removes undesired extrema only in a topological sense, by
creating bridges from undesired extrema towards desired ones, which simply
attaches undesired to desired components via noticeable line artifacts.
    Flattening (d) completely removes components that are associated with
undesired extrema while preserving the shape of desired
components.\vspace{-1em}
  }
  \label{fig_TDAforDifferentSimplificationStrategies}
\end{figure*}

In many applications, datasets produced by acquisition or simulation
currently reach unprecedented levels in terms of size and complexity.
This motivates the design of advanced analysis tools capable of extracting the
relevant information from such datasets, and supporting its
interpretation through interactive visualization and analysis.
This is the purpose of Topological Data Analysis (TDA)~\cite{edelsbrunner2010computational},
which provides a family of generic, robust, and efficient techniques for the
extraction of the inherent structural information in the data.
It has been successfully applied over the last two decades in a number of
visualization and analysis tasks \cite{heine2016survey}.
Popular applications include astrophysics~\cite{sousbie11, shivashankar2016felix},
biological imaging~\cite{carr04, topoAngler, beiBrain18},
chemistry~\cite{harshChemistry, chemistry_vis14, Malgorzata19},
fluid dynamics\julien{~\cite{laney2006understanding, kasten_tvcg11,
bremer_camcs16}},
material sciences~\cite{gyulassy_vis07,gyulassy_vis15,jonas2017,soler_ldav19},
and turbulent combustion~\cite{bremer2011interactive,gyulassy_ev14}.
For scalar data, TDA introduces a number of topological abstractions that capture various types of structural features, such as
the persistence diagram~\cite{Edelsbrunner2002},
the contour tree~\cite{Carr2003, gueunet_tpds19},
the Reeb graph~\cite{reeb1946points, biasotti08, pascucci07},
and the Morse-Smale complex~\cite{Defl15, gyulassy2008practical, robins_pami11, gyulassy_vis18}.

An important aspect of TDA is its ability to provide multi-scale representations of the aforementioned abstractions, which enables users to distinguish noise from features (Figs.~\ref{fig_teaser} and \ref{fig_TDAforDifferentSimplificationStrategies}).
These representations are a critical component in many applications to support multi-scale analysis and visualization.
In particular, several importance measures have been introduced to estimate the
relevance of the features of interest represented by the \emph{critical points} of the scalar field (\autoref{sec_criticalPoints}).
Such measures include topological persistence~\cite{Edelsbrunner2002}
and geometrical measures on level sets~\cite{carr04}.
Moreover, as the notion of a feature of interest greatly depends on the application,
users often characterize the importance of extrema
with ad-hoc importance measures.
For instance, Carr~et~al.~\cite{carr04}
showed that importance measures based on a hyper-volume were often more
effective for medical data than topological persistence, and
Guenther~et~al.~\cite{chemistry_vis14} demonstrated that critical points corresponding to relevant chemical interactions can be isolated by combined thresholding of multiple chemical quantitites.
In general, once an importance measure has been established, multi-scale
representations in TDA can be obtained either in
\emph{(i)}~a~post-processing step (by iteratively simplifying the computed abstractions), or in
\emph{(ii)}~a~pre-processing step (by simplifying the data \emph{before} computing the topological abstractions).
In the first case~\emph{(i)}, a tailored algorithm based on iterative cancellations
must be derived for each \jonasRevision{specific} topological abstraction.
\jonasRevision{
    In the second case~\emph{(ii)}, the values of the original scalar field have
    to be altered in a pre-processing step such that the resulting simplified
scalar
field only exhibits extrema deemed relevant by
some
importance measure
while still being
close
to the original scalar field.
    Although the latter strategy requires a perturbation of the original scalar
    field,
    it
    has the advantage of being generic (since
any importance measure can be used to classify critical points)
    and agnostic (since the simplified scalar field can be used like any other
    scalar field as an input for subsequent TDA).
    This strategy decouples the simplification from the topological
    abstractions themselves,
    which enables software frameworks%
    ---such as the Topology ToolKit~(TTK)~\cite{ttk}---to mutualize and
modularize implementations of different algorithms.
    Specifically, all topological abstractions computed with TTK on the
    simplified field---e.g. critical points~\cite{banchoff1970critical},
merge~\cite{gueunet2017task} and contour trees~\cite{gueunet_tpds19}, Reeb
graphs~\cite{gueunet2019reeb}, and Morse-Smale complexes~\cite{ttk}---benefit
from the same pre-simplification without being aware that the input field was
actually simplified. This is particularly useful in advanced analysis
scenarios
comining
multiple abstractions~\cite{chemistry_vis14,
laney2006understanding}.
    Finally, the pre-simplification of the scalar field does not prohibit further post-simplification of the computed abstractions.
    For these reasons, this paper focuses on the pre-simplification of scalar data.%
}

In addition to the approaches introduced for the persistence-driven
topological simplification of scalar data~\cite{edelsbrunner_socg06, attali_topoinvis09, bauer_dcg12},
Tierny and Pascucci proposed a generic algorithm that supports arbitrary importance measures~\cite{tierny_vis12}.
These simplification algorithms can \jonasRevision{again} be classified
into
two strategies:
bridging (a.k.a. carving) and flattening (a.k.a. flooding).
In short, bridging connects undesired contours to desired
ones
via small bridges \jonasRevision{(technically, along separatrices)}, thus only
removing them in a topological
sense~(\autoref{fig_TDAforDifferentSimplificationStrategies}c).
In contrast, flattening completely removes contours induced by undesired extrema while preserving the shape of desired contours~(\autoref{fig_TDAforDifferentSimplificationStrategies}d).
\jonasRevision{However, both strategies are limited to the removal of
extremum-saddle pairs, which notably excludes saddle-saddle pair removal in
3D~(\autoref{sec_limitations}).}
\julienRevision{Additionally, the two strategies have an impact on feature geometry as they alter the underlying scalar field.
In particular, bridging adds line artifacts between contours, and flattening introduces flat-plateaus that also impact contained integral lines.}
\jonasRevision{But most importantly, despite the acceptable asymptotic complexity (linearithmic time) of both strategies, existing algorithms suffer from performance issues due to their global and sequential nature.}
Specifically, simplification can empirically account for up to 90\% of overall
computation time in many TDA pipelines~(\autoref{sec_interactiveExploration}).

This paper revisits the problem of topological simplification of scalar data to address this performance issue.
Given a scalar field $f$ and a selection of extrema to preserve,
the proposed algorithm creates, based on localized flattening, a function $g$
that only exhibits the selected set of extrema and has a small distance
$||f-g||_\infty$
for data fitting purposes.
Moreover, the proposed localized topological simplification (LTS) algorithm~(\autoref{sec_localizedTopologicalSimplification}) is output-sensitive as it only visits the regions of the data where simplification is needed; resulting in an immediate speedup over previous approaches.
Due to its localized nature, it can be parallelized, and we present---to our knowledge---the first parallel approach to topological simplification of scalar data.
We provide performance benchmarks obtained with our
OpenMP~\cite{dagum1998openmp} implementation that achieves a high parallel
efficiency (\julien{70\%}).
For the
case where
the considered importance measure is topological
persistence,
we present a variant of our framework adapted to persistence-driven
simplification.
Given an input threshold~$\epsilon$, we show how to efficiently
combine LTS with a partial
computation of the persistence diagram.
For applications where a relevant value of~$\epsilon$ can be estimated a priori, this further improves performance.
Extensive experiments on synthetic and real-world data on a
commodity workstation
report an observed speedup of \speedup{36} compared to previous approaches.
This significant improvement is illustrated in various TDA scenarios (\autoref{sec_interactiveExploration}),
where LTS brings the overall analysis time from minutes down to seconds, hence enabling interactive multi-scale exploration of the features present in the data.

\vspace*{1em}
\noindent
\textbf{Contributions}%

        \vspace{0.45em}
        \noindent
        \textbf{1) Localized Topological Simplification (LTS)~(\autoref{sec_localizedTopologicalSimplification})}
        In contrast to state-of-the-art approaches, which visit the entire input domain, LTS only processes the regions of the domain where the data actually needs simplification.
        This makes LTS output sensitive, which is particularly relevant for the problem of topological simplification, as noise frequently corresponds to a small fraction of the input domain.
        In the presented experiments (\autoref{sec_results}), this \jonas{already} yields an average speedup over previous, global approaches between \speedup{3} and \speedup{5}.

        \vspace{0.45em}
        \noindent
        \textbf{2) Parallel Topological Simplification~(\autoref{sec_parallelTopologicalSimplification})}
        LTS can be efficiently parallelized due to its localized nature.
        The presented experiments demonstrate that a shared-memory parallel implementation based on OpenMP yields a high parallel efficiency on real-life datasets (\julien{70\%}).
        On a commodity workstation, this results in an overall speedup of \speedup{36} over previous, sequential~approaches.
        \noindent
        \textbf{3) Parallel Persistence-Driven
Simplification~(\autoref{sec_persistenceBasedSimplification})}
        For the special case where extrema are selected based on topological persistence,
        an adapted version of LTS partially computes the persistence diagram and simultaneously simplifies all features below a given persistence threshold.
        For applications where persistence is a relevant criterion and a threshold is known a priori, this specialized algorithm further improves performances.
        This scenario is particularly relevant for \emph{batch} and \emph{in situ} processing, where noise below a conservative threshold can be removed prior to any analysis.

        \vspace{0.45em}
        \noindent
        \textbf{4) Reference Implementation (additional material)}
        We provide a reference C++ implementation of all presented algorithms
        that can be used to replicate the experiments of \autoref{sec_results}
and to perform benchmarks.

\section{Related Work}
\label{sec_relatedWork}
As described before, topological abstractions can be simplified in a post-process \emph{(i)}
with a tailored algorithm specific to each abstraction.
Such algorithms have been introduced for
the contour tree~\cite{carr04},
Reeb graph~\cite{pascucci07},
and Morse-Smale complex~\cite{gyulassy2008practical}.
Our approach focuses on the \emph{pre}-simplification of data
\emph{(ii)}---i.e., prior to the computation of any
topological abstraction.
Existing methods for data pre-simplification can be classified in the
following two categories.

\begin{figure}[b]
  \begin{center}
    \input{figures/flatteningVsCarving/illustration}
  \end{center}

  \vspace*{-1em}
  \mycaption{
    Differences between removing undesired maxima (gray discs) while preserving
desired maxima (white disc) from an input scalar field $f$ via bridging~(orange)
and flattening~(red).
    Bridging
    creates a monotone path from an undesired maximum along
    separatrices towards a desired maximum by \emph{elevating} the value of the
corresponding vertices (orange chains).
    It modifies the global shape of level sets
    by joining the three super-level set components (dark gray areas above
dashed line) with a connected path (see also
\autoref{fig_TDAforDifferentSimplificationStrategies}).
    In contrast, flattening \emph{lowers} all the vertices
    of an undesired
hill to the closest saddle.
    Hence, flattening completely removes the contours induced by undesired maxima while preserving the shape and topology of other contours.
    Here, only the unmodified super-level set component in the middle would remain.
  }
  \label{fig_IllustrationOfSimplificationStrategies}
\end{figure}

\noindent
\textbf{Numerical methods} simplify an input scalar field given some constraints on the extrema to preserve, by computing a numerically optimized solution.
Such methods usually incorporate the extrema to preserve as hard constraints, while optimizing a geometrical criterion, such as smoothness.
Bremer~et~al.~\cite{bremer2004topological} introduced the first technique in
this line of work in which a simplified Morse-Smale complex
drives a per-cell, iterative simplification of the data via Laplacian
smoothing.
This method requires computing and simplifying the Morse-Smale complex,
which can be computationally expensive.
Moreover,
the method can require many time-intensive Laplacian
iterations to complete.
\jonasRevision{Weinkauf~et~al.~\cite{WeinkaufGS10} extend the work of Bremer~et~al.~\cite{bremer2004topological} by utilizing bi-Laplacian optimization to additionally enforce the continuity of the gradient across the separatrices of the Morse-Smale complex.}
In geometry processing, several techniques have been introduced for the computation of smooth scalar fields given a small number of critical points serving as constraints~\cite{NiGH04, GingoldZ06}.
For instance, Patan\'e~et~al.~\cite{PataneF09} introduce a technique combining least-squares approximation and Tikhonov regularization for the topology-driven simplification of scalar fields.
These approaches can also be extended to 3D scalar
fields~\cite{GuntherJRSSW14}\jonasRevision{ and to some extent to vector
fields~\cite{tricoche2001continuous,zhang2006vector}}.

Although numerical methods can produce smooth outputs that respect given topological constraints,
they have two primary limitations.
First, the iterative nature of the employed solvers
results in long computation times (typically minutes),
which prohibits the interactive exploration of different simplification parameters.
Second,
they are prone to numerical instabilities that can result from an unsuitable
triangulation of the input domain or the numerical sensitivity of the
geometrical operators used in the optimization.
Tierny and Pascucci~\cite{tierny_vis12} illustrate that such instabilities frequently occur during Laplacian optimization and result in the presence of additional spurious critical points that prevent the solution from conforming strictly to the input constraints.

\noindent
\textbf{Combinatorial methods}, in contrast, are designed to generate outputs that are guaranteed by construction to conform to the input constraints.
This category of techniques can be seen as complementary to numerical
approaches, as combinatorial methods can be used to fix, in a post-process, the
possible artifacts generated by numerical methods.
The first approach for the combinatorial simplification of scalar data can be
attributed to Edelsbrunner~et~al. for their work on
persistence-driven simplification
~\cite{edelsbrunner_socg06}.
Given an input function $f$ and its persistence diagram~\cite{Edelsbrunner2002}
(\autoref{sec_persistenceDiagram}), the authors
first simplify the diagram by removing all features below a persistence
threshold $\epsilon$.
In addition to showing that a scalar function $g$ strictly admitting this simplified diagram exists, they also introduce an algorithm to compute such a function $g$ with a bounded error to the input ($||f-g||_\infty \leq \epsilon$).
Their work can be viewed as a generalization of prior works in terrain modeling, where only minimum-saddle persistence pairs were simplified~\cite{soille04, carrPhD, AgarwalAY06}.
Attali~et~al.~\cite{attali_topoinvis09} introduced a simplification algorithm for the case of filtrations of simplicial complexes.
In the context of Discrete Morse theory \cite{forman98},
Bauer~et~al.~\cite{bauer_dcg12} showed that persistence-driven simplification
could be achieved with a combination of
\emph{flattening}
and \emph{bridging} that
results in a minimized error bound $||f-g||_\infty$.
Although these methods were initially
introduced
for two-dimensional domains, they apply
readily for domains of higher dimensions.
However, these methods---as well as our approach---do not support the
removal of saddle-saddle pairs in 3D \julien{(see \autoref{sec_conclusions} for
further discussion)}.

Despite the guaranteed correctness of their outputs,
these combinatorial
methods suffer from several limitations.
First, they all use \emph{bridging},
which introduces undesirable visual line
artifacts in the output, where the bridged integral lines connect simplified
features and therefore often stand out~(\autoref{fig_IllustrationOfSimplificationStrategies},~orange).
This is particularly detrimental if the simplification is used as a pre-process to data segmentation.
As shown in \autoref{fig_TDAforDifferentSimplificationStrategies}c, this can result in the identification of large regions that include the canceled features through a thin connection.
In contrast, \emph{flattening}-based methods completely remove undesired extrema by flattening their corresponding regions to the value of their paired saddle~(\autoref{fig_IllustrationOfSimplificationStrategies},~red).
Therefore, flattening is better suited for data segmentation as simplified features are discarded from post-process segmentations~(\autoref{fig_TDAforDifferentSimplificationStrategies}d).
Second, methods that manipulate filtrations~\cite{attali_topoinvis09} or Discrete Morse functions~\cite{bauer_dcg12} require an extra post-processing step to convert their output into a piecewise linear scalar field (which is the traditional
representation for scalar data). In particular, this step involves the
subdivision of the triangulation (one new vertex per $d$-simplex,
with $d \ge 1$). This can increase the size of the triangulation \julien{up to
an order of magnitude}, which may not be acceptable in certain applications.
Third,
these approaches focus on the special case where the critical points to
cancel are identified according to topological persistence.
As discussed previously, many applications come with their own importance measures.
To address this, Tierny and Pascucci~\cite{tierny_vis12} introduced a
flattening-based approach, directly producing a piecewise linear scalar field on
its output, and which supports the cancellation of an arbitrary selection of
extrema.
However,
\jonasRevision{their} technique follows a \emph{white-list} approach that
removes undesired extrema by globally constraining the sub- and superlevel set
components of the desired extrema.
This results in an intrinsically sequential algorithm, processing the data globally.
In contrast, LTS follows a \emph{black-list} approach by only simplifying the regions that correspond to undesired extrema.
Moreover, the
LTS approach can be efficiently parallelized, which results in a
significant performance gain that enables interactive topological
simplification.

\section{Technical Background}
\label{sec_background}

This section details the technical background of the proposed methodology.
We refer the reader to reference text books~\cite{edelsbrunner2010computational} for a comprehensive introduction to topological data analysis.

\subsection{Critical Points of Piecewise Linear Scalar Fields}
\label{sec_criticalPoints}

The input of the proposed methodology is a piecewise-linear (PL) scalar
field $f:\domain\rightarrow\range$, where real-valued data is given at the
vertices of a PL $d$-manifold $\domain$,
and values inside higher
dimensional simplices are linearly interpolated via barycentric coordinates.
\jonasRevision{In this work, we additionally require that every PL scalar field
is also injective on the vertices of~$\domain$, which can be
enforced for any input scalar field with a symbolic perturbation inspired by \emph{Simulation of
Simplicity}~\cite{edelsbrunner1990simulation}.
\julienRevision{This can be achieved by sorting the list of vertices
$\vertices$ in increasing $f$ order, and enforcing that two consecutive values
are strictly monotonically increasing (by the addition of an arbitrarily small
constant, see \autoref{sec_osf}).}}

The topological features of~$f$ can be tracked with the notion of the so-called \emph{sublevel} set, noted ${\sublevelset{f}(w) = \{ p \in \domain ~ | ~ f(p) < w\}}$, which is defined as the pre-image of the interval $(-\infty, w)$~by~$f$.\vspace*{-0.2em}
Symmetrically, the \emph{superlevel} set is defined as ${\superlevelset{f}(w) = \{ p \in \domain ~ | ~ f(p) > w\}}$.
In the smooth setting, the topology of these sets (in 3D their connected components, cycles, and voids) can only change at specific locations, named the \emph{critical points} of~$f$~\cite{milnor1963morse}.
In the PL setting, Banchoff~\cite{banchoff1970critical} introduced a local
characterization of critical points based on
their \emph{star} and \emph{link}.
The \emph{star}~$\Star(v)$ of a vertex $v \in \domain$ is the set of its co-faces:
${\Star(v) = \{ \simplex \in \domain ~|~ v < \sigma \}}$.
The \emph{link}~$\Link(v)$ of $v$ consists of the faces~$\face$ of the simplices~$\simplex\in\Star(v)$ with an empty intersection with $v$:
${\Link(v) = \{ \face \in \domain ~ | ~ \face < \simplex, ~ \simplex\in \Star(v), ~ \face \cap v = \emptyset\}}$.
The \emph{lower link}~$\lowerlink(v)$ of $v$ consists of the simplices of
$\Link(v)$
lower
than $f(v)$:
${\lowerlink(v) =}$ ${\{ \simplex \in \Link(v) ~ | ~ \forall u \in \sigma, ~ f(u) < f(v)\}}$.
The \emph{upper link} is defined symmetrically:
${\upperlink(v) = \{ \simplex \in \Link(v) ~ | ~ \forall u \in \sigma, ~ f(u) > f(v)\}}$.
The lower and upper links of a vertex $v$ are illustrated in red and orange
in \autoref{fig_cp}.
A vertex $v$ is \emph{regular} (\autoref{fig_cp}a) if and only if both $\lowerlink(v)$ and $\upperlink(v)$ are simply connected and not empty.
For such vertices, the sub- and superlevel sets do not change their topology as they span $\Star(v)$.
Otherwise, $v$ is called a \emph{critical point} of~$f$~\cite{banchoff1970critical}.
These can be classified with regard to their \emph{index} $\Index(v)$,
which is equal to $0$ for local minima ($\lowerlink(v) = \emptyset$, \autoref{fig_cp}b),
to $d$ for local maxima ($\upperlink(v) = \emptyset$, \autoref{fig_cp}c),
and otherwise to $i$ for $i$-saddles ($0 < i < d$, \autoref{fig_cp}d).
Saddles for which the number of connected component of $\lowerlink(v)$ or
$\upperlink(v)$ is greater than $2$ are called \emph{degenerate}.
All other saddles are \emph{simple}.
As discussed in \autoref{sec_localizedTopologicalSimplification}, the proposed algorithm handles degenerate saddles implicitly.

\subsection{General Topological Simplification}
Let $\criticalSet_f$ be the set of critical points of $f$. Tierny and Pascucci
\cite{tierny_vis12} introduce the notion of \emph{general
topological simplification} as follows:\vspace*{-0.25em}

\begin{definition}[General Topological Simplification]
Given a PL scalar field $f : \domain \rightarrow \range$,
a \emph{general topological simplification}
of $f$ is a PL scalar field $g : \domain \rightarrow \range$ such that the
critical points of $g$ form a sub-set of $\criticalSet_f$, i.e., $\criticalSet_g
\subseteq \criticalSet_f$ with identical indices $\Index$ and locations.\vspace*{-0.25em}
\end{definition}

\noindent
In other words, a general topological simplification $g$ is a variant of a
scalar field $f$, from which critical points have been removed. In most
applications,
it is also
desirable that $g$ remains close to the input $f$ (with a small distance
$||f-g||_\infty$) for data fitting purposes.

Tierny and Pasucci~\cite{tierny_vis12}
discuss
the
possible critical point removals.
Based on the analysis of the
connectivity of the sublevel~sets~of~$f$, they show that the removal of a
minimum~$m$ is necessarily accompanied with the removal of a simple saddle~$s$,
where~$\sublevelset{f}$ changes its number of connected components.
Symmetrically, the removal of a maximum induces the removal of a
simple saddle. Thus, they introduce an approach for topological simplification
which is based on the constrained construction of the sublevel sets of~$g$, where
the connectivity of~$\sublevelset{g}$ is globally controlled to
enforce the preservation of minimum and maximum constraints,
respectively noted~$\criticalSet_g^0$~and~$\criticalSet_g^d$.
While this discussion is mostly carried out for 2-dimensional domains, this
approach to extremum removal readily applies to domains of higher
dimensions and has been implemented as a default
simplification mechanism in the \emph{``Topology ToolKit''} open-source library
\cite{ttk}.
Our methodology follows a compatible workflow and drives the simplification by
user specified sets of extrema to maintain or remove.

\subsection{Topological Persistence}
\label{sec_persistenceDiagram}

General topological simplifications are defined for
arbitrary selections of extrema.
In practice, these are often selected based on application dependent importance measures.
Alternatively, \emph{topological persistence}~\cite{Edelsbrunner2002} is often
used as an established, general-purpose importance measure.
The intuition behind persistence consists of assessing the importance of a
critical point, based on the lifetime of its induced feature in
$\superlevelset{f}(w)$ as one continuously decreases the isovalue $w$.
As $w$ decreases, connected components of
$\superlevelset{f}(w)$ appear and merge at the maxima and $2$-saddles of $f$, respectively (\autoref{fig_propagations}a).
The Elder rule~\cite{edelsbrunner2010computational} stipulates that if two
connected
components---created at the maxima $m_0$ and $m_1$ with $f(m_0) < f(m_1)$---meet
at a given $2$-saddle $s$, then the \emph{youngest} of the two components (the
one created last, at $m_0$) \emph{dies} in favor of the \emph{oldest} one
(created at
$m_1$). In this case, a \emph{persistence pair} $\langle s, m_0\rangle$ is created and its \emph{topological persistence} $p$ is given by $p(\langle s, m_0\rangle) = f(m_0) - f(s)$.
Each maximum $m$---with the exception of the global maximum---can be unambiguously
paired following this strategy and can consequently be assigned a persistence
value, noted $p(m)$.
By convention, the global maximum is paired with the global minimum and therefore assigned a persistence equal to the function~range.

Persistence pairs are usually visualized with the persistence diagram~\cite{edelsbrunner2010computational}~(\autoref{fig_propagations}a, right), which embeds each pair $\langle a,b \rangle$ as a point in the 2D plane at
location $\big(f(a), f(b)\big)$.
There, the persistence of the pair can be
readily visualized as the height of the point to the diagonal. In particular,
features with a high persistence stand out by being far away from the diagonal (dark colored pairs),
while noisy features are typically located in the vicinity of the diagonal (light colored pairs).
The population of persistence pairs is also often visualized with the
\emph{persistence curve}~(\autoref{sec_results}), which plots, for an
increasing threshold~$\epsilon$, the number of pairs more persistent
than~$\epsilon$.
As described above, $2$-saddle/maximum pairs characterize the lifetime of the
connected components of $\superlevelset{f}(w)$. The symmetric reasoning
can be applied in 3D to characterize, with minimum/$1$-saddle pairs, the life
time of its
voids, while the $1$-saddle/$2$-saddle pairs characterize its independent
cycles.
Then, given a threshold $\epsilon$, the notion of
persistence-driven simplification
is a special
case of general simplification, such that
$\forall m \in \criticalSet_g^0~:~p(m) \geq \epsilon$
and
$\forall m\in \criticalSet_g^d~:~p(m) \geq \epsilon$.

\begin{figure}
    \begin{center}
        \newcommand{\starLinkExampleXScale}{0.78}
\newcommand{\starLinkExampleYScale}{0.6}

\begin{tikzpicture}[xscale=\starLinkExampleXScale,yscale=\starLinkExampleYScale]

    \newcommand{\starLinkExampleRegNodes}{
        \node[highlightedSimplexVertexB] (D) at (-1,0.5) {+};
        \node[highlightedSimplexVertexB] (E) at (0,1) {+};
        \node[highlightedSimplexVertexB] (F) at (1,0.7) {+};
        \node[simplexVertex] (H) at (0,0) {\textbf{v}};
        \node[highlightedSimplexVertexR] (I) at (-0.6,-0.8) {\rule{0.425em}{0.075em}};
        \node[highlightedSimplexVertexR] (J) at (0.7,-0.6) {\rule{0.425em}{0.075em}};
    }

    \starLinkExampleRegNodes

    \fill [simplexTriangle] (D.center) -- (E.center) -- (H.center) -- cycle;
    \fill [simplexTriangle] (D.center) -- (H.center) -- (I.center) -- cycle;
    \fill [simplexTriangle] (E.center) -- (F.center) -- (H.center) -- cycle;
    \fill [simplexTriangle] (F.center) -- (J.center) -- (H.center) -- cycle;
    \fill [simplexTriangle] (H.center) -- (I.center) -- (J.center) -- cycle;

    \draw[highlightedSimplexEdgeB] (D)--(E)--(F);
    \draw[highlightedSimplexEdgeR] (I)--(J);

    \starLinkExampleRegNodes

    \node at (0,-1.7) {\hspace*{-3em}a) Regular Point\hspace*{-3em}};
\end{tikzpicture}%
\hfill%
\begin{tikzpicture}[xscale=\starLinkExampleXScale,yscale=\starLinkExampleYScale]
    \newcommand{\starLinkExampleMinNodes}{
        \node[highlightedSimplexVertexB] (D) at (-1,0.5) {+};
        \node[highlightedSimplexVertexB] (E) at (0,1) {+};
        \node[highlightedSimplexVertexB] (F) at (1,0.7) {+};
        \node[simplexVertex] (H) at (0,0) {\textbf{v}};
        \node[highlightedSimplexVertexB] (I) at (-0.6,-0.8) {+};
        \node[highlightedSimplexVertexB] (J) at (0.7,-0.6) {+};
    }

    \starLinkExampleMinNodes

    \fill [simplexTriangle] (D.center) -- (E.center) -- (H.center) -- cycle;
    \fill [simplexTriangle] (D.center) -- (H.center) -- (I.center) -- cycle;
    \fill [simplexTriangle] (E.center) -- (F.center) -- (H.center) -- cycle;
    \fill [simplexTriangle] (F.center) -- (J.center) -- (H.center) -- cycle;
    \fill [simplexTriangle] (H.center) -- (I.center) -- (J.center) -- cycle;

    \draw[highlightedSimplexEdgeB] (D)--(E)--(F)--(J)--(I)--(D);

    \starLinkExampleMinNodes

    \node at (0,-1.7) {\hspace*{-3em}b) Minimum\hspace*{-3em}};
\end{tikzpicture}%
\hfill%
\begin{tikzpicture}[xscale=\starLinkExampleXScale,yscale=\starLinkExampleYScale]
    \newcommand{\starLinkExampleMaxNodes}{
        \node[highlightedSimplexVertexR] (D) at (-1,0.5) {\rule{0.425em}{0.075em}};
        \node[highlightedSimplexVertexR] (E) at (0,1) {\rule{0.425em}{0.075em}};
        \node[highlightedSimplexVertexR] (F) at (1,0.7) {\rule{0.425em}{0.075em}};
        \node[simplexVertex] (H) at (0,0) {\textbf{v}};
        \node[highlightedSimplexVertexR] (I) at (-0.6,-0.8) {\rule{0.425em}{0.075em}};
        \node[highlightedSimplexVertexR] (J) at (0.7,-0.6) {\rule{0.425em}{0.075em}};
    }

    \starLinkExampleMaxNodes

    \fill [simplexTriangle] (D.center) -- (E.center) -- (H.center) -- cycle;
    \fill [simplexTriangle] (D.center) -- (H.center) -- (I.center) -- cycle;
    \fill [simplexTriangle] (E.center) -- (F.center) -- (H.center) -- cycle;
    \fill [simplexTriangle] (F.center) -- (J.center) -- (H.center) -- cycle;
    \fill [simplexTriangle] (H.center) -- (I.center) -- (J.center) -- cycle;

    \draw[highlightedSimplexEdgeR] (D)--(E)--(F)--(J)--(I)--(D);

    \starLinkExampleMaxNodes

    \node at (0,-1.7) {\hspace*{-3em}c) Maximum\hspace*{-3em}};
\end{tikzpicture}%
\hfill%
\begin{tikzpicture}[xscale=\starLinkExampleXScale,yscale=\starLinkExampleYScale]
    \newcommand{\starLinkExampleSadNodes}{
        \node[highlightedSimplexVertexB] (D) at (-1,0.5) {+};
        \node[highlightedSimplexVertexB] (E) at (0,1) {+};
        \node[highlightedSimplexVertexR] (F) at (1,0.7) {\rule{0.425em}{0.075em}};
        \node[simplexVertex] (H) at (0,0) {\textbf{v}};
        \node[highlightedSimplexVertexR] (I) at (-0.6,-0.8) {\rule{0.425em}{0.075em}};
        \node[highlightedSimplexVertexB] (J) at (0.7,-0.6) {+};
    }

    \starLinkExampleSadNodes

    \fill [simplexTriangle] (D.center) -- (E.center) -- (H.center) -- cycle;
    \fill [simplexTriangle] (D.center) -- (H.center) -- (I.center) -- cycle;
    \fill [simplexTriangle] (E.center) -- (F.center) -- (H.center) -- cycle;
    \fill [simplexTriangle] (F.center) -- (J.center) -- (H.center) -- cycle;
    \fill [simplexTriangle] (H.center) -- (I.center) -- (J.center) -- cycle;

    \draw[highlightedSimplexEdgeB] (D)--(E);

    \starLinkExampleSadNodes

    \node at (0,-1.7) {\hspace*{-3em}d) Saddle Point\hspace*{-3em}};
\end{tikzpicture}
    \end{center}
    \vspace*{-1.25em}
    \mycaption{%
        Vertex classification:
        if the lower and upper link of a vertex $v$ (red and orange) are both connected, then $v$ is regular~(a).
        It is a minimum or maximum if it has no lower or upper link, respectively~(b-c).
        If its lower or upper link consists of multiple components, then $v$ is called a saddle~(d).\vspace*{-1em}%
    }
    \label{fig_cp}
\end{figure}

\begin{figure*}
    \vspace{-1ex}
\input{figures/exampleDefinitions.tex}

\hspace*{-1.75em}
\begin{tikzpicture}[xscale=\tptsOneDscaleX,yscale=\tptsOneDscaleY]
    \begin{scope}
        \examplePersistencePairs
    \end{scope}

    \begin{scope}[xshift=52em]
        \exampleArbitraryPropagations
    \end{scope}

    \begin{scope}[xshift=108em]
        \examplePersistenceBasedPropagations
    \end{scope}

    \draw[filtered, dashed] (16.6,-1.5) -- ++(0,8);
    \draw[filtered, dashed] (34.2,-1.5) -- ++(0,8);
\end{tikzpicture}
    \vspace*{-1.05em}

    \mycaption{
        Critical point pairs and their corresponding domain segments (a)
computed with superlevel set propagations (b) and
persistence-driven
propagations (c).
        Each maximum (disc), its paired saddle (diamond), and
its corresponding domain segment (background) are shown with the same color.
        In the first step of LTS (b), superlevel set propagations determine---optionally in parallel---for each undesired maximum (dashed discs) its corresponding saddle and domain segment~(Secs.~\ref{sec_oneHill}-\ref{sec_parallelTopologicalSimplification}).
        In the special case of
        persistence-driven simplification (c), non-persistent maxima are
identified by initiating for all maxima superlevel set propagations that
terminate as soon as they exceed a given persistence threshold (dashed lines).
        Thus, maxima paired with a saddle (dashed discs) do not exceed the persistence threshold and can subsequently be simplified~(\autoref{sec_persistenceBasedSimplification}).\vspace*{-1.5em}
    }
    \label{fig_propagations}
\end{figure*}

\section{Localized Topological Simplification (LTS)}
\label{sec_localizedTopologicalSimplification}
Given a PL scalar field~$f$ and a subset of its minima~$\preserve{f}{0}$~and~maxima~$\preserve{f}{d}$,
the proposed algorithm derives via localized flattening a function~$g$ that
has a small deviation $||f-g||_\infty$ and
only exhibits the specified~extrema:\vspace*{-0.25em}
\begin{equation}
    \criticalSet_{g}^{0} = \preserve{f}{0} \subseteq \criticalSet_{f}^{0}
    \text{\hspace*{1.25em}and\hspace*{1.25em}}
    \criticalSet_{g}^{d} = \preserve{f}{d} \subseteq \criticalSet_{f}^{d}.\vspace*{-0.25em}
    \label{eq_constraints123}
\end{equation}
The following description focuses on the removal of maxima, which can be imagined as the process of \mbox{\emph{\hspace*{0.0em}``flattening hills''}} in a terrain
(\autoref{fig_IllustrationOfSimplificationStrategies},~red).
Minima are processed
symmetrically.
\subsection{Order-Based Representation}
\label{sec_osf}
The critical point characterization presented in \autoref{sec_criticalPoints}
classifies the vertices of $\domain$ as regular or critical
without any ambiguity as long as every vertex has a distinct value from all its
neighbors, i.e., if there exists a non-ambiguous global vertex order.
In this case, every link can be binarily partitioned into \emph{lower} and \emph{upper} link components~(\autoref{fig_cp}).
Since this property does not hold for every PL scalar field $f : \domain \rightarrow \range$,
the proposed algorithm first derives based on $f$ an intermediate scalar field
$\order{f}$ that satisfies this condition via a specific variant of
\textit{Simulation of Simplicity}~\cite{edelsbrunner1990simulation}.
Next, the algorithm simplifies $\order{f}$ to $\order{g}$ based on the same
extrema conditions as defined for~$f$, and finally simplifies $f$~to~$g$ in a post-processing step based on~$\order{g}$.
The overall process is summarized in the inset commutative diagram.
\begin{wrapfigure}{l}{0.5\linewidth}
    \begin{center}
    \vspace*{-2em}
        \begin{tikzcd}[row sep=small]
            \centering
              f : \domain \rightarrow \range \arrow[r, dashrightarrow] \arrow[d]
            & g: \domain \rightarrow \range \\
              \order{f} : \domain \rightarrow \Natural \arrow[r]
            & \order{g} : \domain \rightarrow \Natural \arrow[u]
        \end{tikzcd}
    \vspace*{-2em}
    \end{center}
\end{wrapfigure}
Let $\vertices$
be the list of
vertices of~$\domain$,
sorted by increasing~$f$~values,
where vertices with the same scalar value are
disambiguated based on their original position in memory.
Then, $\order{f} : \domain \rightarrow \Natural$ is the so-called \emph{order field} of~$f$ that maps each vertex~$v\in\domain$ to its position~$\order{f}(v)$ in~$\vertices$.
By construction,
\emph{(i)} $\order{f}$ is injective on the vertices~of~$\domain$, and \emph{(ii)} the
critical points~of~$\order{f}$ are a \emph{superset} of the critical points
of~$f$.
The first property \emph{(i)} guarantees that now every vertex has a distinct
value from its neighbors, which can thus be classified without any
ambiguity.
The second property \emph{(ii)} is essential for topological simplification,
which basically boils down to reordering vertices such that the resulting new order field~$\order{g}$
respects
\autoref{eq_constraints123}.
To this end, note that every critical point of $f$ is also a critical point of $\order{f}$, but $\order{f}$ might also have additional critical points that result form the disambiguation.
These points, however, will be implicitly removed by LTS (as
they do not belong to the set of extrema constraints
$\preserve{\order{f}}{0}=\preserve{f}{0}$~and~$\preserve{\order{f}}{d}=\preserve
{f}{d}$), which will yield
a new
order field $\order{g}$ whose critical points respect
\autoref{eq_constraints123}.

Finally,
$f$ can be
turned into $g$ in a post-process  by enforcing that the
numerical values of $g$ are monotonically \julienRevision{(and injectively)}
increasing with $\order{g}$.
This post-process monotonicity enforcement results in the characteristic
\emph{hill flattening} observed~with~flattening-based simplification.
\subsection{Removal of an Individual Maximum}
\label{sec_oneHill}

This section first details the proposed approach in a simple configuration where only one maximum $m$ must be removed,
i.e.,\vspace*{-0.65em}
\begin{equation}
    \criticalSet^0_{\order{g}}=\preserve{\order{f}}{0}=\criticalSet^0_{\order{f}}
    \quad\text{ and }\quad
    \criticalSet^d_{\order{g}}=\preserve{\order{f}}{d}=\criticalSet^d_{\order{f}}\setminus\{m\}.\vspace*{-0.65em}
    \label{eq_singleMaxConstraint}
\end{equation}
To this end, the following procedure will iteratively transform the order
field~$\order{f}$~to~$\order{g}$ in four steps.
\autoref{fig_localSimplification} provides a running example,
where the task is to remove the maximum~$m$, with initial value
~$\order{f}(m)=22$.

\vspace*{0.3em}
\noindent
\textbf{1) Superlevel Set Component Computation}
As discussed in \autoref{sec_background},
when continuously decreasing an isovalue~$w$,
a
connected component of $\superlevelset{\order{f}}(w)$ is created at the local
maximum $m$ of $\order{f}$.
As $w$ further decreases, this component eventually merges with another component at a saddle~$s$.
Let $\superlevelsetcomponent{s}{m}$ denote the connected component of $\superlevelset{\order{f}}\big(\order{f}(s)\big)$ containing $m$.
The vertex~set~$\superlevelsetcomponent{s}{m}$ can be computed via a so-called
\emph{superlevel set propagation} that initializes
$\superlevelsetcomponent{s}{m}$ with $m$, and then iteratively adds the largest
neighbor~of $\superlevelsetcomponent{s}{m}$~to~$\superlevelsetcomponent{s}{m}$
until a vertex $v$ with a larger, unvisited neighbor~$n$ is added, i.e., $\order{f}(n) >
\order{f}(v)$ and $n\notin \superlevelsetcomponent{s}{m}$.
This implies that $n$ belongs to a distinct superlevel set component, and
therefore $v$ has to be a saddle (noted~$s$~above).
For convenience $s$ is considered to be an element of
$\superlevelsetcomponent{s}{m}$ in the remaining.
\emph{Sublevel set propagations} are defined symmetrically.
In \autoref{fig_localSimplification}a, vertices are
added to \jonasRevision{$\superlevelsetcomponent{\emptyset}{22}$} (arrows, orange
triangles) until a vertex ($10$) with a higher unvisited neighbor ($23$) is
found. This vertex corresponds to the paired saddle $s$ of $m$.

\noindent
\textbf{2) Localized Simplification}
\label{sec_method_ls}
To remove the maximum $m$
while preserving all other maxima,
the vertices of $\superlevelsetcomponent{s}{m}$ need to be rearranged in a new global order $\order{g}$ such that vertices outside $\superlevelsetcomponent{s}{m}$ preserve their
old order (\autoref{eq_localConstraints1}), $s$ is the only maximum of $\order{g}$ restricted to
$\superlevelsetcomponent{s}{m}$ (\autoref{eq_localConstraints2}),
and all minima of $\order{g}|\superlevelsetcomponent{s}{m}$ are vertices with neighbors outside $\superlevelsetcomponent{s}{m}$ (\autoref{eq_localConstraints3}):\vspace*{-0.5em}
\begin{alignat}{3}
    \forall u\in\domain\text{ and }\forall v\in\domain\setminus \superlevelsetcomponent{s}{m} &\;\;\;:\;\;&&
        \order{f}(u)<\order{f}(v)
        \rightarrow
        \order{g}(u)<\order{g}(v)\label{eq_localConstraints1}\\
    \criticalSet_{\order{g|\superlevelsetcomponent{s}{m}}}^{d} &\;\;=\;\;&& \{s\}\label{eq_localConstraints2}\\
    \forall v\in\criticalSet_{\order{g|\superlevelsetcomponent{s}{m}}}^{0}
&\;\;\;:\;\;&& \exists\,
n\,\in\,Lk(v),
~n \notin \superlevelsetcomponent{s}{m}
\label{eq_localConstraints3}
\end{alignat}

\vspace*{-0.5em}
\noindent
As detailed in Step~3, if $\order{g}$ satisfies
Eqs.~\ref{eq_localConstraints1}-\ref{eq_localConstraints3}, then $\order{g}$ also
satisfies \autoref{eq_singleMaxConstraint}.
A key insight is that enforcing these conditions on~$\order{g|\superlevelsetcomponent{s}{m}}$ is itself a special case of topological simplification, localized to vertex values of~$\superlevelsetcomponent{s}{m}$.
To simplify notations, let $\order{l_0}:\superlevelsetcomponent{s}{m}\rightarrow\Natural$ denote the local order of vertices of $\superlevelsetcomponent{s}{m}$ (values of opaque vertices in \autoref{fig_localSimplification}b) induced by the initial global order field~$\order{f}$ (vertex values in~\autoref{fig_localSimplification}a).

This localized simplification problem can be solved by adapting the iterative simplification algorithm of Tierny and Pasucci~\cite{tierny_vis12} to this special setting.
To summarize, their algorithm takes as input constraints an explicit list of \emph{authorized} extrema to preserve.
Then, the algorithm alternates passes to remove \emph{unauthorized} maxima and
minima by respectively constraining the connectivity of the super- and sublevel
set components of the authorized extrema.
However, a maxima pass might introduce additional unauthorized minima that need to be removed with an additional minima pass, and vice versa.
The authors show that iteratively alternating between
minimum and maximum
passes is guaranteed to converge to a global output order that only
exhibits the authorized extrema.

However, in the localized setting, the notion of authorized extremum does not
readily apply as $l_0$ does not necessarily exhibit other extrema than $m$.
Thus, LTS first introduces an authorized maximum at
$s$ by setting its corresponding value to
infinity, and then introduces an authorized minimum at some vertex
$v\in\superlevelsetcomponent{s}{m}$ different from $s$ that admits a neighbor
outside $\superlevelsetcomponent{s}{m}$ by setting its corresponding value to
negative infinity, i.e., ${\order{l_0}(s)\hookleftarrow
+\infty}$~and~${\order{l_0}(v)\hookleftarrow
-\infty}$~(\autoref{fig_localSimplification}b).
During all iterations, LTS always preserves $s$ as the only authorized maximum,
and dynamically
updates
the set of authorized minima
with the vertices which
are minima
on $\superlevelsetcomponent{s}{m}$ \emph{and} admit neighbors outside
$\superlevelsetcomponent{s}{m}$.

In the first iteration, the algorithm performs on $\superlevelsetcomponent{s}{m}$ a superlevel set propagation initiated at
the only authorized maximum (arrows of \autoref{fig_localSimplification}b).
The inverse order in which vertices are added during this propagation is guaranteed to yield only one maximum: $s$.
Thus, this inverse order becomes the new local~order~$l_1$ on $\superlevelsetcomponent{s}{m}$~(\autoref{fig_localSimplification}c).

Next, if unauthorized minima are present---i.e., minima that do not admit neighbors outside $\superlevelsetcomponent{s}{m}$---the algorithm initiates symmetrically a
sublevel set propagation from all current authorized minima.
In \autoref{fig_localSimplification}c, such a propagation is initiated at the authorized minimum~$00$ to remove the unauthorized minima~$01$~and~$02$.
The order in which vertices are added during this propagation is guaranteed to omit the unauthorized minima, and therefore becomes the new local order~$l_2$~(\autoref{fig_localSimplification}d).
Then, LTS alternates between maxima and minima iterations until the last computed local order~$\order{l}$ only exhibits authorized extrema~(\autoref{fig_localSimplification}e).

\noindent
\textbf{3) Local to Global Order}
\label{sec_method_ltgo}
At the end of the previous step, the last computed local order
$\order{l}$ is
guaranteed to have only one maximum which is located at $s$, and all its minima are located next to a vertex outside $\superlevelsetcomponent{s}{m}$.
At this point, it is necessary to update the global order $\order{f}$ to
$\order{g}$ by reordering the vertices of $\superlevelsetcomponent{s}{m}$ in the
global order such that $s$ preserves its old position, and all other vertices of
$\superlevelsetcomponent{s}{m}$ form a
contiguous segment, located immediately before $s$, and sorted by $\order{l}$.
The position of each vertex of $\domain$ in the resulting new global order
yields the new order field $\order{g}$ that~satisfies~
Eqs.~\ref{eq_localConstraints1}-\ref{eq_localConstraints3}, as detailed next.

Since $s$ was originally a saddle there has to exist a neighbor
${n\in\domain\setminus\superlevelsetcomponent{s}{m}}$ of $s$ with
${\order{f}(s)<\order{f}(n)}$, so from \autoref{eq_localConstraints1} follows that
${\order{g}(s)<\order{g}(n)}$, which makes $s$ no longer a maximum on
$\domain$.
Similarly, every minimum
${v\in\criticalSet_{\order{g|\superlevelsetcomponent{s}{m}}}^{0}}$ has a neighbor
${n\in\domain\setminus\superlevelsetcomponent{s}{m}}$ where
${\order{g}(n)<\order{g}(v)}$---this follows from the fact that
${\superlevelsetcomponent{s}{m}}$ was computed by a superlevel set
propagation---which makes $v$ no longer a minimum on entire $\domain$.
Hence, all extrema of ${\order{g}|\superlevelsetcomponent{s}{m}}$ are not extrema
of~${\order{g}|\domain}$, and all extrema outside $\superlevelsetcomponent{s}{m}$
are preserved.
Note that $\superlevelsetcomponent{s}{m}$ may have contained minima that also
got removed during the local simplification (as illustrated in
\autoref{fig_localSimplification}); a known artifact of flattening-based
simplification.
If these minima are required to be preserved, their existence can be enforced in an optional post-process that makes them smaller than all their neighbors.
It follows that $\order{g}$ satisfies \autoref{eq_singleMaxConstraint} as
required, so the maximum $m$ has truly been removed combinatorially.

\vspace*{0.5em}
\noindent
\textbf{4) Symbolic to Numerical Perturbation}
\label{sec_method_stnp}
The previous process only changes the order of vertices such that the maximum $m$ of $\order{f}$ is no longer classified as an extremum of $\order{g}$, while preserving all other extrema.
This symbolic perturbation is sufficient for most TDA pipelines since the
criticality of vertices only depends on their order (\autoref{sec_osf}).
In certain cases, however, it may be useful to reflect the new order in the
original numerical values by deriving a new scalar field $g$ based on
$\order{g}$.
To this end, $g$ is first initialized to $f$, and then every vertex of $\domain$ is visited in decreasing value of $\order{g}$.
If during this process a visited vertex $v$ has a higher $g$ value than its predecessor $v'$ (i.e., $g(v) >
g(v')$), then its $g$ value is updated to be
smaller than its predecessor (i.e., $g(v)
\leftarrow g(v') - \zeta$, with $\zeta$ arbitrarily small)
which effectively flattens the hill $\superlevelsetcomponent{s}{m}$.
This also guarantees that $g$ only deviates from $f$ on $\superlevelsetcomponent{s}{m}$ with at most the height of the hill, i.e., $||f-g||_\infty\;\leq\;f(m)-f(s)$.

\begin{figure}
    \vspace*{-1.5em}
    \begin{center}
        \input{figures/localSimplification.tex}
    \end{center}
    \vspace*{-0.5em}

    \mycaption{
        Illustration of the localized simplification process, where
        authorized (solid) and unauthorized (dotted) maxima, minima, and
        saddles are shown with orange, red, and gray nodes, respectively.
        Here, the algorithm is tasked to update the global order $\order{f}$ by removing the maximum~$m$ with index~${\order{f}(m)=22}$~(a).
        The core concept of LTS is that removing the maximum~$m$ corresponds to
        reordering the vertices of its superlevel set
component~$\superlevelsetcomponent{s}{m}$~(orange triangles, the order is
denoted by the arrows between the vertices) in a new
local order~$\order{l}:\superlevelsetcomponent{s}{m}\rightarrow\Natural$ such
that $s$ is the only maximum in $\superlevelsetcomponent{s}{m}$, and all minima
in $\superlevelsetcomponent{s}{m}$ have a neighbor outside
$\superlevelsetcomponent{s}{m}$.
        This can be enforced by iteratively removing unauthorized extrema locally, via
        alternating superlevel and sublevel set propagations~(b-d) until only authorized extrema remain in $\superlevelsetcomponent{s}{m}$~(e).
        Finally, the new local order $\order{l}$ is used to derive a new global order $\order{g}$ such that $\superlevelsetcomponent{s}{m}$ no longer contains any extrema~(f).
        Note, removing a maximum $m$ also removes minima inside its corresponding superlevel set component $\superlevelsetcomponent{s}{m}$ (e.g., vertices $12$ and $13$ (a)); a known artifact of flattening-based simplification.
        If desired, their existence can be enforced in an optional post-process that lowers their order until they are identified as minima again.
        \vspace*{-1.5em}
    }
    \label{fig_localSimplification}
\end{figure}

\subsection{Removal of Multiple Maxima}
\vspace*{-0.25em}
\label{sec_method_rmm}

In principle, the strategy described in \autoref{sec_oneHill} can be used iteratively to remove multiple maxima one by one.
However, regions corresponding to discarded maxima can form localized clusters,
i.e., \emph{hill chains}.
Since hills in these chains are nested, flattening each hill one after the other
with this strategy results in multiple passes over the same portions of
the data.
To address this issue, it is necessary to slightly modify
the first three steps of the localized
simplification process:
Step~1 needs to compute the combined region of every hill chain,
Step~2 needs to locally simplify each of these combined regions, and
Step~3 needs to integrate all local orders into the global order.

To this end, Step~1 has to initiate from each undesired maximum ${m_i \in
\criticalSet^d_{\order{f}}\setminus\preserve{\order{f}}{d}}$\vspace*{-0.1em} a superlevel set
propagation $\superlevelsetcomponent{s_i}{m_i}$ (\autoref{sec_oneHill}).
\jonasRevision{When a propagation reaches a saddle $s_i$, then the propagation stops if at least one higher saddle neighbor has not been visited so far by any propagation (\autoref{fig_propagations}b: light propagations), or otherwise---i.e., if now all higher saddle neighbors have been visited by some propagation---the current propagation merges with the other propagations that reached the same saddle and then continues towards the next saddle (\autoref{fig_propagations}b: dark propagations).}
In the latter case, this means at an algorithmic
level that the priority queues used for the propagations which stopped at $s_i$
need to be merged with that of the current propagation to guarantee the valid
extraction of the superlevel set component.
As suggested by Gueunet et al.~\cite{gueunet_tpds19} in the context of computing augmented contour trees,
LTS uses Fibonacci heaps~\cite{fredman1987fibonacci,cormen} to model the propagation priority queues since they support constant time merge operations; guaranteeing an overall linearithmic time complexity.
At the end of Step~1, LTS extracts one superlevel set component per hill chain
(\autoref{fig_propagations}b), which can be flattened at once in Step~2.
The list of vertices of each hill chain can be efficiently composed by
maintaining a Union-Find data structure~\cite{cormen}.

\jonasRevision{Next, Step~2 independently computes a local order for every hill
chain as described in \autoref{sec_oneHill}, and then Step~3 iteratively
integrates every computed local order into the global vertex order as described
before.
Finally, the 
global procedure that computes a\julienRevision{n
injective} numerical perturbation
remains identical (\autoref{sec_oneHill}).}

\subsection{Parallel Removal of Multiple Maxima}
\label{sec_parallelTopologicalSimplification}

This subsection describes a shared-memory
parallelization of LTS.
Step~1 can be trivially parallelized on a per
discarded maximum basis. However, propagations need to be synchronized at
saddles,
to discover which propagation is the last one to reach a
saddle $s$.
This can be implemented with an atomic counter that records the
number of remaining unvisited higher neighbors, which only reaches zero for the
last propagation visiting $s$.
Step~2 can be trivially parallelized on a per region basis.
Step~3 can be parallelized by computing an order
${\order{i}:\domain\rightarrow\Natural \times
\Natural}$ operating along
the initial order $\order{f}$
(disambiguating distinct regions) and the local orders
$\order{l}$:
\vspace*{-1.5em}
\begin{equation}
    \order{i}(v)=
    \begin{cases}
        \big(\;\order{f}(s)\,,\,+\infty\;\big)&\text{if $v=s$ for some $\superlevelsetcomponent{s}{m}$,}\\
        \big(\;\order{f}(s)\,,\,\order{l}(v)\;\big)&\text{if $v\neq s \wedge v\in\superlevelsetcomponent{s}{m}$ for some $\superlevelsetcomponent{s}{m}$, and}\\
        \big(\;\order{f}(v)\,,\,\;\;0\;\;\;\big)&\text{otherwise.}
    \end{cases}
\end{equation}

\vspace*{-0.75em}
\noindent
Sorting all vertices based on $\order{i}$ with a parallel sorting
algorithm---such as GNU parallel sort~\cite{singler2008gnu}---yields the new
global order $\order{g}$.
Finally, Step~4 cannot be parallelized in its current form as it requires a
sequential pass over all vertices. However, this step induces negligible
computation times in practice, which does not impair parallel performances.
\subsection{Persistence-Driven Simplification}
\label{sec_persistenceBasedSimplification}

In the special case where all maxima below a given persistence threshold
$\epsilon$
need to be removed, an adapted version of LTS first detects all non-persistent
maxima during the superlevel set propagations of Step~1,
and then locally simplifies their corresponding regions as described before.
Specifically, Step~1 now has to initiate, from \emph{each} maximum~$m$, a superlevel set propagation that now also tracks the scalar difference between~$m$ and its last visited vertex.
If this difference exceeds~$\epsilon$, then the current propagation immediately terminates, since it must correspond to a \emph{persistent maximum} (\autoref{fig_propagations}c, dark orange and dark red).
All other propagations---i.e., propagations that merged with persistent propagations~(\autoref{fig_propagations}c, light red) or propagations that terminated at a saddle with unvisited larger neighbors~(\autoref{fig_propagations}c, light orange and light gray)---must correspond to \emph{non-persistent maxima}.
As illustrated in \autoref{fig_propagations}c, this procedure partially computes the persistence diagram by only constructing critical point pairs
with a persistence smaller than~$\epsilon$.

This process can be parallelized as described in
\autoref{sec_parallelTopologicalSimplification}, and the remaining steps are
identical to previous descriptions. In particular, only the regions
corresponding to non-persistent maxima are processed by the local flattening
procedure. At the end of this process, since each region is flattened by a
height equal to the function difference between its highest maximum and its
lowest saddle (\autoref{sec_oneHill}), the output function $g$ guarantees that $||f-g||_\infty \leq \epsilon$ and that all maxima less persistent than $\epsilon$ have indeed been removed.

\newcommand{\mytimes}{\text{$\times$}}

\begin{table}[t]
    \fontsize{7}{7}\selectfont
    \def\arraystretch{0.5}
    \setlength{\tabcolsep}{0.57em}

    \begin{tabular}{|l|r||r||r|r||r|r|r|}
    \hline
    \multirow{2}{*}{Dataset}
      & \multicolumn{1}{c||}{$|\domain|$}
      & \multicolumn{1}{c||}{BL}
      & \multicolumn{2}{c||}{LTS (1 core)}
      & \multicolumn{3}{c|}{LTS (12 cores)}\\
      & \multicolumn{1}{c||}{}
      & \multicolumn{1}{c||}{Time}
      & \multicolumn{1}{c|}{Time}
      & \multicolumn{1}{c||}{\textbf{S.U.}}
      & \multicolumn{1}{c|}{Time}
      & \multicolumn{1}{c|}{\textbf{S.U.}}
      & \multicolumn{1}{c|}{\emph{P.E.}}\\
    \hhline{|=|=||=||=|=||=|=|=|}
      Silicium &
      $0.1 \mytimes 10^{6}$ &
      0.100 &
      0.034 &
      \textbf{2.9} &
      0.004 &
      \textbf{25.0} &
      \emph{71 \%}\\
      \arrayrulecolor{lightgray}\hline
      Cells &
      $0.8 \mytimes10^{6}$ &
      1.410 &
      0.303 &
      \textbf{4.7} &
      0.031 &
      \textbf{45.5} &
      \emph{81 \%}\\
      \arrayrulecolor{lightgray}\hline
      OceanVortices &
      $1 \mytimes10^{6}$ &
      1.321 &
      0.437 &
      \textbf{3.0} &
      0.049 &
      \textbf{27.0} &
      \emph{74 \%}\\
      \arrayrulecolor{lightgray}\hline
      Foot &
      $17 \mytimes10^{6}$ &
      20.505 &
      6.673 &
      \textbf{3.1} &
      0.785 &
      \textbf{26.1} &
      \emph{71 \%}\\
      \arrayrulecolor{lightgray}\hline
      Random &
      $17 \mytimes10^{6}$ &
      65.201 &
      8.178 &
      \textbf{8.0} &
      0.927 &
      \textbf{70.3} &
      \emph{74 \%}\\
      \arrayrulecolor{lightgray}\hline
      Turbulence &
      $17 \mytimes10^{6}$ &
      53.444 &
      12.015 &
      \textbf{4.4} &
      1.255 &
      \textbf{42.6} &
      \emph{80 \%}\\
      \arrayrulecolor{lightgray}\hline
      Backpack &
      $98 \mytimes10^{6}$ &
      174.599 &
      67.572 &
      \textbf{2.6} &
      7.454 &
      \textbf{23.4} &
      \emph{76 \%}\\
      \arrayrulecolor{lightgray}\hline
      Jet &
      $134 \mytimes10^{6}$ &
      167.879 &
      52.121 &
      \textbf{3.2} &
      6.366 &
      \textbf{26.4} &
      \emph{68 \%}\\
      \arrayrulecolor{black}\hline
      \cellcolor[HTML]{e1e1e1}Silicium &
      \cellcolor[HTML]{e1e1e1} $0.1 \mytimes10^{6}$ &
      \cellcolor[HTML]{e1e1e1} 0.111 &
      \cellcolor[HTML]{e1e1e1} 0.051 &
      \cellcolor[HTML]{e1e1e1} \textbf{2.2} &
      \cellcolor[HTML]{e1e1e1} 0.011 &
      \cellcolor[HTML]{e1e1e1} \textbf{10.1} &
      \cellcolor[HTML]{e1e1e1} \emph{39 \%}\\
      \arrayrulecolor{lightgray}\hline
      \cellcolor[HTML]{e1e1e1}Cells &
      \cellcolor[HTML]{e1e1e1}$0.8 \mytimes10^{6}$ &
      \cellcolor[HTML]{e1e1e1} 1.767 &
      \cellcolor[HTML]{e1e1e1} 1.339 &
      \cellcolor[HTML]{e1e1e1} \textbf{1.3} &
      \cellcolor[HTML]{e1e1e1} 0.410 &
      \cellcolor[HTML]{e1e1e1} \textbf{4.3} &
      \cellcolor[HTML]{e1e1e1} \emph{27 \%}\\
      \arrayrulecolor{lightgray}\hline
      \cellcolor[HTML]{e1e1e1}OceanVortices &
      \cellcolor[HTML]{e1e1e1}$1 \mytimes10^{6}$ &
      \cellcolor[HTML]{e1e1e1} 1.118 &
      \cellcolor[HTML]{e1e1e1} 1.562 &
      \cellcolor[HTML]{e1e1e1} \textbf{0.7} &
      \cellcolor[HTML]{e1e1e1} 1.193 &
      \cellcolor[HTML]{e1e1e1} \textbf{0.9} &
      \cellcolor[HTML]{e1e1e1} \emph{11 \%}\\
      \arrayrulecolor{lightgray}\hline
      \cellcolor[HTML]{e1e1e1}Foot &
      \cellcolor[HTML]{e1e1e1}$17 \mytimes10^{6}$ &
      \cellcolor[HTML]{e1e1e1} 19.498 &
      \cellcolor[HTML]{e1e1e1} 11.171 &
      \cellcolor[HTML]{e1e1e1} \textbf{1.7} &
      \cellcolor[HTML]{e1e1e1} 2.037 &
      \cellcolor[HTML]{e1e1e1} \textbf{9.6} &
      \cellcolor[HTML]{e1e1e1} \emph{46 \%}\\
      \arrayrulecolor{lightgray}\hline
      \cellcolor[HTML]{e1e1e1}Random &
      \cellcolor[HTML]{e1e1e1}$17 \mytimes10^{6}$ &
      \cellcolor[HTML]{e1e1e1} 51.576 &
      \cellcolor[HTML]{e1e1e1} 28.363 &
      \cellcolor[HTML]{e1e1e1} \textbf{1.8} &
      \cellcolor[HTML]{e1e1e1} 6.684 &
      \cellcolor[HTML]{e1e1e1} \textbf{7.7} &
      \cellcolor[HTML]{e1e1e1} \emph{35 \%}\\
      \arrayrulecolor{lightgray}\hline
      \cellcolor[HTML]{e1e1e1}Turbulence &
      \cellcolor[HTML]{e1e1e1}$17 \mytimes10^{6}$ &
      \cellcolor[HTML]{e1e1e1} 49.075 &
      \cellcolor[HTML]{e1e1e1} 18.435 &
      \cellcolor[HTML]{e1e1e1} \textbf{2.7} &
      \cellcolor[HTML]{e1e1e1} 4.232 &
      \cellcolor[HTML]{e1e1e1} \textbf{11.6} &
      \cellcolor[HTML]{e1e1e1} \emph{36 \%}\\
      \arrayrulecolor{lightgray}\hline
      \cellcolor[HTML]{e1e1e1}Backpack &
      \cellcolor[HTML]{e1e1e1}$98 \mytimes10^{6}$ &
      \cellcolor[HTML]{e1e1e1} 169.415 &
      \cellcolor[HTML]{e1e1e1} 94.319 &
      \cellcolor[HTML]{e1e1e1} \textbf{1.8} &
      \cellcolor[HTML]{e1e1e1} 28.655 &
      \cellcolor[HTML]{e1e1e1} \textbf{5.9} &
      \cellcolor[HTML]{e1e1e1} \emph{27 \%}\\
      \arrayrulecolor{lightgray}\hline
      \cellcolor[HTML]{e1e1e1}Jet &
      \cellcolor[HTML]{e1e1e1} $134 \mytimes10^{6}$&
      \cellcolor[HTML]{e1e1e1} 165.160 &
      \cellcolor[HTML]{e1e1e1} 57.871 &
      \cellcolor[HTML]{e1e1e1} \textbf{2.9} &
      \cellcolor[HTML]{e1e1e1} 7.748 &
      \cellcolor[HTML]{e1e1e1} \textbf{21.3} &
      \cellcolor[HTML]{e1e1e1} \emph{62 \%}\\
      \arrayrulecolor{lightgray}\hline
    \arrayrulecolor{black}\hline
    \end{tabular}

  \vspace*{0.8em}

    \caption{Performance comparison of the baseline
approach~\mbox{\cite{tierny_vis12, ttk}}~(BL) versus the sequential and parallel execution of LTS for realistic
extremum selections (white lines: 1\% of the function range) and a stress
case
(gray lines: only global extrema are preserved). Timings are in seconds,
speedup (\emph{S.U.}) is relative to the baseline approach, and parallel
efficiency (\emph{P.E.}) is defined as parallel speedup divided by the number of cores.\vspace*{-0.1em}
}
    \label{tab:performanceLTS}
\end{table}

\begin{table}[t]
    \fontsize{7}{7}\selectfont
    \def\arraystretch{0.5}
    \setlength{\tabcolsep}{0.405em}

    \centering

    \begin{tabular}{|l|r||r||r||r|r||r|r|}
    \hline
    \multirow{2}{*}{Dataset}
      & \multicolumn{1}{c||}{$|\domain|$}
      & \multicolumn{1}{c||}{Diagram}
      & \multicolumn{1}{c||}{Dia. + BL}
      & \multicolumn{2}{c||}{Dia. + LTS}
      & \multicolumn{2}{c|}{Pers.-LTS}\\
      & \multicolumn{1}{c||}{}
      & \multicolumn{1}{c||}{Time}
      & \multicolumn{1}{c||}{Time}
      & \multicolumn{1}{c|}{Time}
      & \multicolumn{1}{c||}{\textbf{S.U.}}
      & \multicolumn{1}{c|}{Time}
      & \multicolumn{1}{c|}{\textbf{S.U.}}\\
    \hhline{|=|=||=||=||=|=||=|=|}
      Silicium &
      $0.1 \mytimes 10^{6}$ &
      \emph{0.055} &
      0.156 &
      0.060 &
      \textbf{2.6} &
      0.010 &
      \textbf{15.6} \\
      \arrayrulecolor{lightgray}\hline
      Cells &
$0.8 \mytimes10^{6}$ &
      \emph{0.113} &
      1.505 &
      0.145 &
      \textbf{10.4} &
      0.049 &
      \textbf{30.7} \\
      \arrayrulecolor{lightgray}\hline
      OceanVortices &
$1 \mytimes10^{6}$ &
      \emph{0.208} &
      1.530 &
      0.258 &
      \textbf{5.9} &
      0.132 &
      \textbf{11.6} \\
      \arrayrulecolor{lightgray}\hline
      Foot &
$17 \mytimes10^{6}$ &
      \emph{2.355} &
      23.391 &
      3.207 &
      \textbf{7.3} &
      1.080 &
      \textbf{21.7} \\
      \arrayrulecolor{lightgray}\hline
      Random &
$17 \mytimes10^{6}$ &
      \emph{14.808} &
      81.039 &
      16.440 &
      \textbf{4.9} &
      2.853 &
      \textbf{28.4} \\
      \arrayrulecolor{lightgray}\hline
      Turbulence &
$17 \mytimes10^{6}$ &
      \emph{2.703} &
      67.232 &
      4.024 &
      \textbf{16.7} &
      2.313 &
      \textbf{29.1} \\
      \arrayrulecolor{lightgray}\hline
      Backpack &
$98 \mytimes10^{6}$ &
      \emph{24.397} &
      253.386 &
      32.061 &
      \textbf{7.9} &
      9.364 &
      \textbf{27.1} \\
      \arrayrulecolor{lightgray}\hline
      Jet &
$134 \mytimes10^{6}$ &
      \emph{149.327} &
      316.305 &
      155.827 &
      \textbf{2.0} &
      7.558 &
      \textbf{41.9} \\
      \arrayrulecolor{black}\hline

      \cellcolor[HTML]{e1e1e1}Silicium &
\cellcolor[HTML]{e1e1e1} $0.1 \mytimes10^{6}$ &
      \cellcolor[HTML]{e1e1e1} \emph{0.055} &
      \cellcolor[HTML]{e1e1e1} 0.155 &
      \cellcolor[HTML]{e1e1e1} 0.066 &
      \cellcolor[HTML]{e1e1e1} \textbf{2.3} &
      \cellcolor[HTML]{e1e1e1} 0.016 &
      \cellcolor[HTML]{e1e1e1} \textbf{9.7} \\
      \arrayrulecolor{lightgray}\hline
      \cellcolor[HTML]{e1e1e1}Cells &
\cellcolor[HTML]{e1e1e1}$0.8 \mytimes10^{6}$ &
      \cellcolor[HTML]{e1e1e1} \emph{0.111} &
      \cellcolor[HTML]{e1e1e1} 1.874 &
      \cellcolor[HTML]{e1e1e1} 0.520 &
      \cellcolor[HTML]{e1e1e1} \textbf{3.6} &
      \cellcolor[HTML]{e1e1e1} 0.456 &
      \cellcolor[HTML]{e1e1e1} \textbf{4.1} \\
      \arrayrulecolor{lightgray}\hline
      \cellcolor[HTML]{e1e1e1}OceanVortices &
\cellcolor[HTML]{e1e1e1}$1 \mytimes10^{6}$ &
      \cellcolor[HTML]{e1e1e1} \emph{0.219} &
      \cellcolor[HTML]{e1e1e1} 1.337 &
      \cellcolor[HTML]{e1e1e1} 1.472 &
      \cellcolor[HTML]{e1e1e1} \textbf{0.9} &
      \cellcolor[HTML]{e1e1e1} 1.224 &
      \cellcolor[HTML]{e1e1e1} \textbf{1.1} \\
      \arrayrulecolor{lightgray}\hline
      \cellcolor[HTML]{e1e1e1}Foot &
\cellcolor[HTML]{e1e1e1}$17 \mytimes10^{6}$ &
      \cellcolor[HTML]{e1e1e1} \emph{2.363} &
      \cellcolor[HTML]{e1e1e1} 21.576 &
      \cellcolor[HTML]{e1e1e1} 4.433 &
      \cellcolor[HTML]{e1e1e1} \textbf{4.9} &
      \cellcolor[HTML]{e1e1e1} 2.596 &
      \cellcolor[HTML]{e1e1e1} \textbf{8.3} \\
      \arrayrulecolor{lightgray}\hline
      \cellcolor[HTML]{e1e1e1}Random &
\cellcolor[HTML]{e1e1e1}$17 \mytimes10^{6}$ &
      \cellcolor[HTML]{e1e1e1} \emph{14.891} &
      \cellcolor[HTML]{e1e1e1} 65.486 &
      \cellcolor[HTML]{e1e1e1} 21.903 &
      \cellcolor[HTML]{e1e1e1} \textbf{3.0} &
      \cellcolor[HTML]{e1e1e1} 9.849 &
      \cellcolor[HTML]{e1e1e1} \textbf{6.6} \\
      \arrayrulecolor{lightgray}\hline
      \cellcolor[HTML]{e1e1e1}Turbulence &
\cellcolor[HTML]{e1e1e1}$17 \mytimes10^{6}$ &
      \cellcolor[HTML]{e1e1e1} \emph{2.725} &
      \cellcolor[HTML]{e1e1e1} 51.907 &
      \cellcolor[HTML]{e1e1e1} 6.853 &
      \cellcolor[HTML]{e1e1e1} \textbf{7.6} &
      \cellcolor[HTML]{e1e1e1} 6.032 &
      \cellcolor[HTML]{e1e1e1} \textbf{8.6} \\
      \arrayrulecolor{lightgray}\hline
      \cellcolor[HTML]{e1e1e1}Backpack &
\cellcolor[HTML]{e1e1e1}$98 \mytimes10^{6}$ &
      \cellcolor[HTML]{e1e1e1} \emph{24.018} &
      \cellcolor[HTML]{e1e1e1} 228.711 &
      \cellcolor[HTML]{e1e1e1} 52.906 &
      \cellcolor[HTML]{e1e1e1} \textbf{4.3} &
      \cellcolor[HTML]{e1e1e1} 19.511 &
      \cellcolor[HTML]{e1e1e1} \textbf{11.7} \\
      \arrayrulecolor{lightgray}\hline
      \cellcolor[HTML]{e1e1e1}Jet &
\cellcolor[HTML]{e1e1e1} $134 \mytimes10^{6}$&
      \cellcolor[HTML]{e1e1e1} \emph{148.811} &
      \cellcolor[HTML]{e1e1e1} 322.559 &
      \cellcolor[HTML]{e1e1e1} 156.641 &
      \cellcolor[HTML]{e1e1e1} \textbf{2.1} &
      \cellcolor[HTML]{e1e1e1} 11.083 &
      \cellcolor[HTML]{e1e1e1} \textbf{29.1} \\
      \arrayrulecolor{lightgray}\hline
    \arrayrulecolor{black}\hline
    \end{tabular}

  \vspace*{0.8em}

  \caption{Performance comparison of persistence-driven simplification
(white lines: 1\% of the function range, gray lines: only global extrema are
preserved) by computing the persistence diagram~\cite{gueunet_tpds19,
ttk}~(Dia, 12~cores) followed either by the baseline (BL)
approach~\cite{tierny_vis12, ttk} or LTS~(12~cores), or by alternatively computing the persistence-driven specialization of LTS (Pers.-LTS, 12~cores).
Speedup (\emph{S.U.}) is relative to \emph{``Dia + BL''}.\vspace*{-1.9em}
  }
  \label{table_persistenceSimplification}
\end{table}

\section{Results}
\label{sec_results}

This section presents experimental results of a C++ implementation of LTS---in
the form of a Topology ToolKit~(TTK)~\cite{ttk} module---obtained on a
desktop computer
with a Xeon CPU (2.6 GHz, 2x6 cores) and with 64 GB of RAM.
The presented datasets have been downloaded from public repositories~\cite{openSciVisDataSets, ttkData}, and the used TDA pipelines consist of modules readily available in TTK.

\subsection{Time Performance}
\label{sec_timePerformance}
As discussed in \autoref{sec_oneHill}, LTS extends the baseline
approach by Tierny and Pascucci~\cite{tierny_vis12} to the local simplification
of sub- and superlevel set components. Thus, LTS admits
the same asymptotic time complexity: $\mathcal{O}\big(N_I
\times |\vertices|log(|\vertices|)\big)$, where $|\vertices|$ is the number of
vertices in $\domain$, and $N_I$ represents the number of iterations of
the algorithm~(\autoref{sec_oneHill}) with $N_I = |\vertices|$ in the
worst case~\cite{tierny_vis12}.
However, in comparison to the baseline approach,
LTS improves run times in three ways.
First, since LTS is localized, \jonasRevision{computational intensive procedures are constraint onto (often small) subsets of the domain.}
In the presented experiments, only up to 2\% of the domain is
\julienRevision{processed}
for realistic
levels of
simplification (white lines in \autoref{tab:performanceLTS}), and up to \julien{70\%} for the most aggressive levels (gray lines).
\jonasRevision{Second, LTS relaxes the constraints on the extrema located on the boundary of sub- and superlevel set components~(\autoref{sec_oneHill}), which has the positive effect that the iterative process converges faster to valid local orders.}
For a given dataset, the maximum number of iterations, over all components to simplify, was rarely above $1$ for realistic simplification levels.
For the most aggressive simplification levels (gray lines in \autoref{tab:performanceLTS}),
only a few components require at most $6$ iterations, while the vast majority of components require only $1$
iteration; resulting in an average number of $1$ iteration for all datasets across all simplification levels.
\jonasRevision{The final speedup results from the parallel processing of individual regions.}
\jonasRevision{These three effects combined---i.e., the parallel simplification of local regions with faster convergence---significantly improve run time performances.
In particular, they make
\julienRevision{the main parts of}
LTS output-sensitive where run time is now a function over the number of
extrema to remove.}

\autoref{tab:performanceLTS} provides a detailed run time comparison between
the baseline approach~\cite{tierny_vis12} and LTS for various datasets.
\jonasRevision{All timings are based on the implementations available in TTK~\cite{ttk}.}
First, for realistic extremum selections (with a
persistence threshold of $1\%$ of the function range), LTS provides an average
speedup in sequential of \speedup{4} over the baseline approach. As discussed above,
this speedup can be explained by the
localized nature of LTS and its faster convergence.
The most important performance gains
can be observed when running LTS in parallel, with an overall
average speedup of \speedup{36} over the baseline approach
(white lines, \autoref{tab:performanceLTS}). The average parallel efficiency
for realistic simplifications (white lines, \autoref{tab:performanceLTS}) is
slightly over $74\%$, which illustrates the good scaling of LTS in this setup.
Second, when stressing LTS with aggressive simplification levels (by
only keeping
\julienRevision{the global minimum and the global maximum,}
gray lines), performances decrease, further
illustrating the output sensitive aspect of LTS.
\julienRevision{The}
average speedup in parallel over the baseline drops down to \speedup{9}, while the
parallel efficiency of LTS drops down to $35\%$. This can be
explained by
\julienRevision{a}
work load imbalance between the threads, since for
aggressive thresholds, some sub- and superlevel set components can become
significantly larger than others, and require more iterations than others.

\begin{figure}[p!]
    \centering
    \adjustbox{width=.9\linewidth,center}{

\newcommand{\tickDelta}{0.05}
\newcommand{\plotScaleX}{0.0775}
\newcommand{\plotScaleY}{0.04}
\newcommand{\plot}[2]{

    \draw [#1, line width=4pt] plot coordinates {#2};


}

\begin{tikzpicture}[xscale=\plotScaleX,yscale=\plotScaleY]
    \foreach \x in {10,20,...,100}{
        \draw[thin,filtered] (\x,0) -- (\x,70);
    }
    \foreach \x in {10,20,...,70}{
        \draw[thin,filtered] (0,\x) -- (100,\x);
    }

    \plot{colorTheme_20}{(0,7.71390000000000000000) (5,8.40790000000000000000) (10,8.99260000000000000000) (15,9.52730000000000000000) (20,10.05470000000000000000) (25,10.59120000000000000000) (30,11.15020000000000000000) (35,11.74000000000000000000) (40,12.35770000000000000000) (45,13.01720000000000000000) (50,13.71520000000000000000) (55,14.37780000000000000000) (60,15.12700000000000000000) (65,15.91160000000000000000) (70,16.73730000000000000000) (75,17.70860000000000000000) (80,18.79620000000000000000) (85,20.03500000000000000000) (90,21.55230000000000000000) (95,23.56760000000000000000) (100,27.95040000000000000000)}

    \plot{colorTheme_10}{(0,62.66736000000000000000) (5,62.61492000000000000000) (10,61.42897000000000000000) (15,62.31372000000000000000) (20,62.20313000000000000000) (25,60.72506000000000000000) (30,61.58750000000000000000) (35,60.86754000000000000000) (40,59.59833000000000000000) (45,59.52322000000000000000) (50,59.48716000000000000000) (55,59.32074000000000000000) (60,58.10074000000000000000) (65,57.66847000000000000000) (70,57.58220000000000000000) (75,56.92229000000000000000) (80,56.69351000000000000000) (85,54.93186000000000000000) (90,53.85107000000000000000) (95,52.51029000000000000000) (100,49.95989000000000000000)}

    \begin{scope}[yshift=4em]
        \plot{colorTheme_00}{(0,.87130000000000000000) (5,.92500000000000000000) (10,.99750000000000000000) (15,1.11490000000000000000) (20,1.18590000000000000000) (25,1.29610000000000000000) (30,1.32940000000000000000) (35,1.47800000000000000000) (40,1.58070000000000000000) (45,1.67770000000000000000) (50,1.72610000000000000000) (55,1.80480000000000000000) (60,1.88570000000000000000) (65,2.03410000000000000000) (70,2.15150000000000000000) (75,2.26500000000000000000) (80,2.40030000000000000000) (85,2.42740000000000000000) (90,2.55480000000000000000) (95,2.84950000000000000000) (100,3.76300000000000000000)}
    \end{scope}

    \draw[thick,arrow,-{Stealth[scale=1]}] (0,0) -- (100,0);
    \draw[thick,arrow,-{Stealth[scale=1]}] (0,0) -- (0,70);

    \node at (-1,-4) {0};

    \foreach \x in {10,20,...,90}{
        \draw[thick] (\x,-\tickDelta/\plotScaleY) -- (\x,\tickDelta/\plotScaleY);
        \node at (\x,-4) {\x};
    }

    \foreach \x in {10,20,...,60}{
        \draw[thick] (\tickDelta/\plotScaleX,\x) -- (-\tickDelta/\plotScaleX,\x);
        \node[anchor=east] at (0,\x) {\x};
    }

    \filldraw[draw=black,fill=white] (10,20) rectangle ++(40,30);

    \draw[line width=5pt, colorTheme_10] (12,45) -- ++(10,0);
    \node[anchor=west] at (22,45) {Baseline~\cite{tierny_vis12}};

    \draw[line width=5pt, colorTheme_20] (12,35) -- ++(10,0);
    \node[anchor=west] at (22,35) {\normalsize LTS-1};

    \draw[line width=5pt, colorTheme_00] (12,25) -- ++(10,0);
    \node[anchor=west] at (22,25) {LTS-12};

    \node at (50,-12) {Percentage of Removed Extrema (\%)};
    \node[rotate=90] at (-9,35) {Simplification Time (s)};

\end{tikzpicture}
    }
    \vspace*{-1.25em}
    \mycaption{Computation time as a function of the percentage of extrema to
    simplify, with a random selection of extrema on the random dataset (gray:
    baseline~\cite{tierny_vis12}, orange: sequential LTS, and red: LTS with 12
cores).}
    \label{fig_randomSimplification}
\end{figure}

\begin{figure}[p!]
    \centering
    \vspace{-1ex}
    \adjustbox{width=\linewidth,center}{
    \includegraphics[
    width=0.495\linewidth]{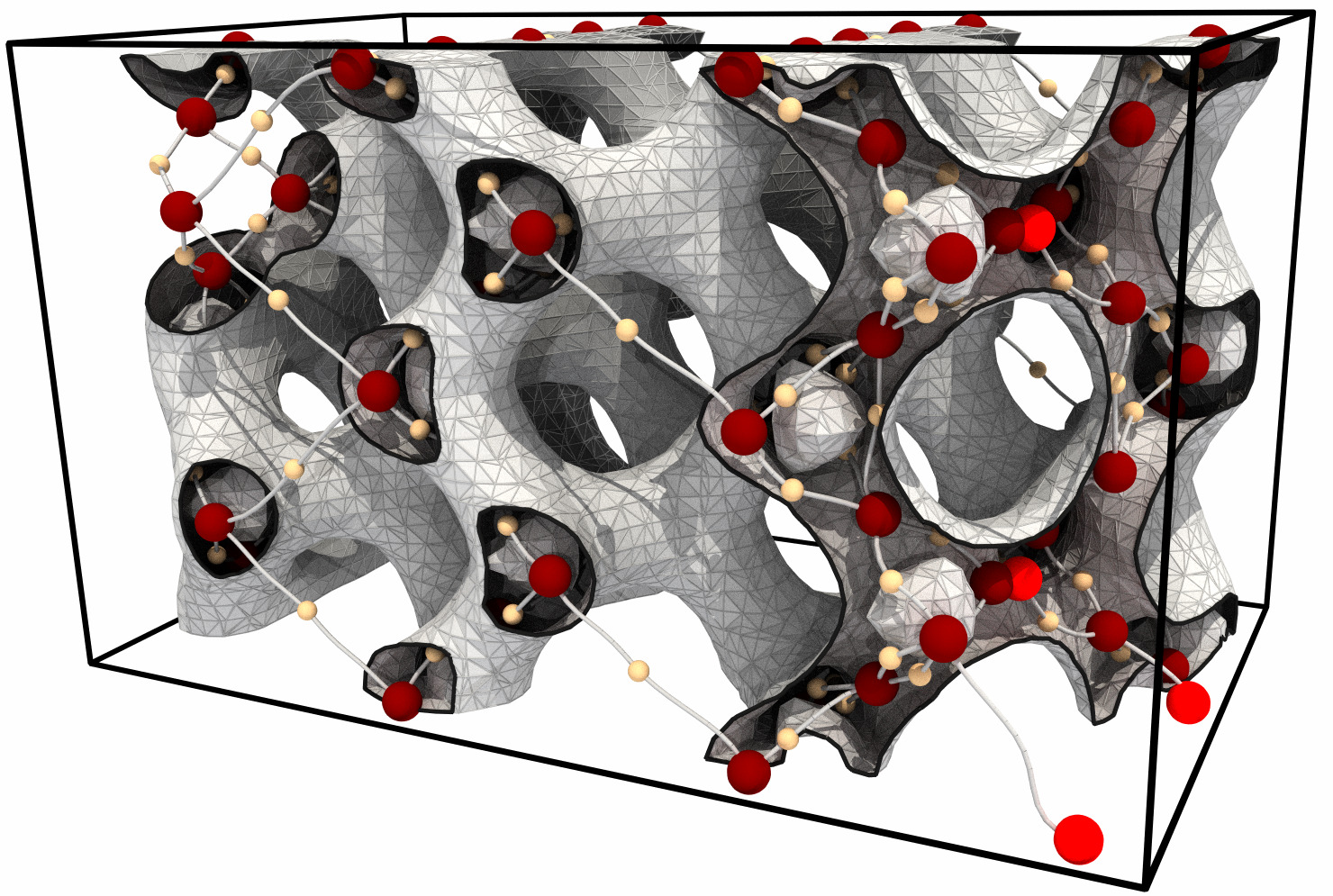}
    \hfill
    \includegraphics[
    width=0.495\linewidth]{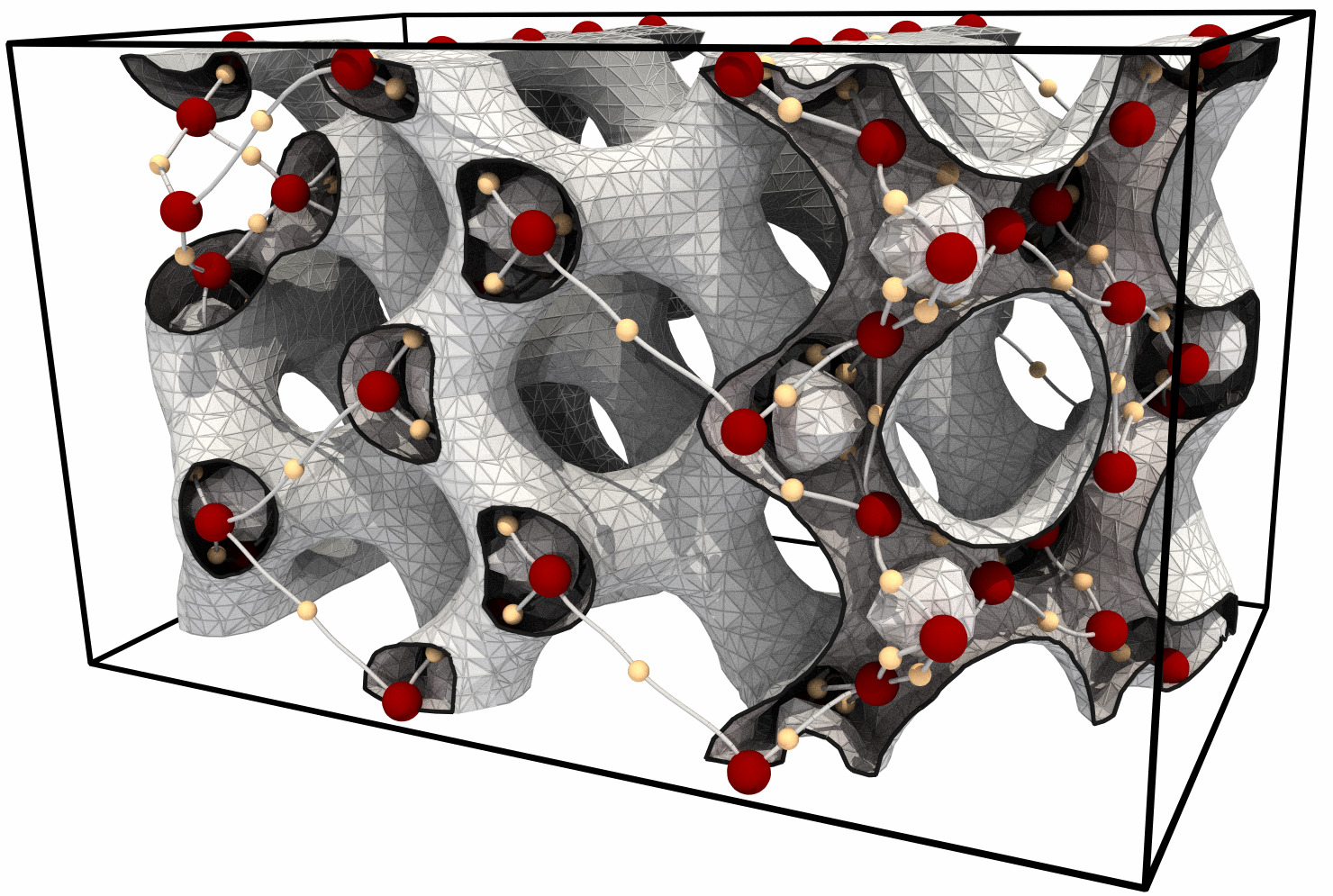}
    }
    \vspace{-3ex}
    \mycaption{Persistence-driven simplification of
    \julienRevision{maxima not corresponding to silicium
atoms~\cite{openSciVisDataSets}}
    (left, bright red spheres), prior to the extraction of the correct lattice
    structure via the Morse-Smale complex (dark red:~maxima, orange:
$2$-saddles, white curves: separatrices).
    Computing the persistence diagram~\cite{gueunet_tpds19} and pre-simplifying
the
    data with the baseline approach~\cite{tierny_vis12} takes $0.15s$, while
    LTS requires $0.02s$.
    \vspace*{-2ex}
    }
    \label{fig_silicium}
\end{figure}

\begin{figure}[p!]
    \adjustbox{width=\linewidth,center}{\input{./figures/foot/foot.tex}}
    \vspace*{-1.5em}
    \mycaption{
    Extractions
    \jonasRevision{of bones in a CT scan of a human
foot~\cite{openSciVisDataSets}
    (volume rendering, left), where bones are identified as the regions that
correspond to the leaf-arcs of the
    merge tree. For a persistence
    simplification threshold of~$180$, the bones of the $5$~toes are precisely
    extracted (center),
    \julienRevision{and}
    are further subdivided along the joints at smaller
thresholds ($150$, right). Pre-simplifying the data with the baseline
    approach~\cite{tierny_vis12} takes~$19.6s$, and computing the merge tree
    segmentation~\cite{gueunet_tpds19} requires~$1.8s$. LTS requires
    $1.3s$ to pre-simplify the data, resulting in a \emph{pipeline-speedup} of
\speedup{7}.}}
    \label{fig_foot}
\end{figure}

\begin{figure}[p!]
    \adjustbox{width=\linewidth,center}{\input{./figures/cells/cells.tex}}
    \vspace*{-1.5em}
    \mycaption{Extracted cells and their nuclei in a microscopy
image~\cite{cellImage} via the Morse-Smale
    complex separatrices (black) and sublevel sets (red). Without simplification
(top left), the analysis suffers from over-segmentation with numerous false
positives. The persistence curve
    (top right) exhibits a clear plateau, indicating a stable
    simplification range separating noise from features, where the appropriate
    persistence threshold still needs to be adjusted interactively.
    False positives are still identified for the left extremity of the plateau
(bottom left),
    while a correct extraction is obtained at the right extremity (bottom
right).
    Pre-simplifying the data with the baseline approach~\cite{tierny_vis12}
takes $1.58s$, whereas LTS only requires $0.18s$.
    Computing the Morse-Smale complex and the sublevel sets takes~$0.94s$.
    \vspace{-1em}
    }
    \label{fig_cell2}
\end{figure}

Table \ref{table_persistenceSimplification} provides a detailed run time
comparison about persistence-driven simplification via the computation of the persistence diagram~\cite{gueunet_tpds19} followed either by the baseline approach~\cite{tierny_vis12} or LTS (12 cores), or alternatively via the persistence-driven specialization of LTS (12
cores, \autoref{sec_persistenceBasedSimplification}).
When used in conjunction with a standard persistence diagram computation (\emph{``Dia. + LTS''}), LTS provides an average speedup of \speedup{7} overall for realistic levels
($1\%$ of the function range, white lines). Since LTS only partially computes
the persistence diagram, the persistence-driven specialization of LTS (right
column)
achieves significant performance gains, with an average speedup of \speedup{26} over the
baseline approach.
Again, for more aggressive simplification levels (gray lines), performances start to deteriorate
as regions which may not
need to be later simplified still need to be visited by the algorithm in order
to compute the partial diagram. Despite this, LTS still provides an
order of magnitude speedup on average.

\jonasRevision{The output-sensitive behavior of LTS is further illustrated in
\autoref{fig_randomSimplification}, which plots the simplification time of the
random dataset as a function over the percentage of randomly selected extrema to remove.
The run time of the baseline approach~\cite{tierny_vis12}~(gray curve)
continuously decreases from $60$ to $50$ seconds, which indicates that the
increasing size of the simplified regions is beneficial to this global approach.
Conversely, the run time of the sequential (orange curve) and parallel (red
curve) execution of LTS is progressively increasing, which indicates that small
regions are beneficial to the local approach.}

\subsection{Application to Interactive Exploration}
\label{sec_interactiveExploration}

\begin{figure*}
  \vspace{-1.5ex}
  \adjustbox{width=\linewidth,center}{
    \input{figures/backpack/backpack.tex}
  }

  \mycaption{
  Level set based feature extraction in the
CT-scan of a backpack (top left). The initial level set counts hundreds of
connected components (640).
\julienRevision{The persistence curve exhibits several kinks, which challenges
the identification of a clear simplification threshold, hence motivating
interactive exploration.}
For aggressive levels, only the denser objects (metal
elements) are maintained in the level set (red objects, top right). For less
aggressive values (bottom, from right to left), less dense objects
(bottles, cords, tools, boxes, etc.) progressively appear in the
segmentation\julienRevision{, in decreasing order of the persistence of the
corresponding topological feature}.
For the leftmost image, the front bottle (dashed black line)
corresponds to
a low-persistence feature whose maximum has been selected geometrically.
For each simplification, LTS computes within  seconds
($13.567s$, leftmost image) while state-of-the-art
techniques~\cite{tierny_vis12} take minutes to simplify ($179.175s$), resulting
in an overall \emph{pipeline-speedup} of \speedup{12}.
    \vspace*{-1em}
}
  \label{fig_backpack}
\end{figure*}

The significant speedup of LTS enables the interactive exploration of simplification parameters and their effect on TDA.
Figures \ref{fig_silicium}--\ref{fig_backpack} present different use cases of interactive exploration
scenarios, where features of acquired and simulated datasets are characterized based on level sets, merge trees, and Morse-Smale complexes.
The context of the analysis and the details of the topological pipeline used
after simplification is reported in the caption of each figure.

\autoref{fig_silicium} illustrates the interest of persistence-driven
simplification, which can be employed in batch mode to remove, in a
pre-processing, features below a conservative persistence level.
While topological simplification is nearly always needed to cope with noise, the relevant amount of
simplification is often not known a priori and therefore needs to be adjusted interactively on a
per dataset basis.
Moreover, there may be more than one relevant simplification level, and visualizing the entire hierarchy is often of interest for analysts.
This motivates efficient algorithms capable of supporting interactive exploration sessions.
In the use cases presented in Figs.~\ref{fig_cell2}--\ref{fig_backpack},
the persistence diagram and its persistence curve are computed in a pre-process.
Next, the user interactively explores different potentially interesting simplification levels,
\julienRevision{typically}
reported by the persistence curve. After each modification of the simplification
level, the TDA pipeline under consideration is
recomputed on the simplified data.
For all of these scenarios, pre-simplifying the data with the
baseline approach~\cite{tierny_vis12} took longer than running the
rest of the TDA pipeline. For the larger datasets,
this  pre-simplification is the main bottleneck, representing up to
90\% of the computation.
In contrast, LTS provides for the simplification step alone speedups of an order of magnitude, making it possible to update the analysis within seconds while the state-of-the-art needs several~minutes.

\subsection{Limitations}
\label{sec_limitations}

In contrast to the baseline approach by Tierny and Pascucci~\cite{tierny_vis12},
LTS uses a black-list strategy and processes only the regions which need simplification.
To keep track of these regions, heavier data structures need to be maintained through the propagations (Fibonacci heaps/Union-Find), but as demonstrated in~\autoref{sec_timePerformance}, LTS still provides superior performances, even in sequential.
The parallel performance of LTS depends significantly on the work load balance among the threads.
Specifically, the number of tasks---i.e., propagations and local simplifications---decreases over time, and at some point the algorithm processes less tasks than available cores.
Thus, parallel efficiency starts to degrade
when
this time interval takes up larger fractions of the total computation time.
However, for the realistic simplification scenarios presented in \autoref{sec_timePerformance}, the work load among threads is well balanced.

\jonasRevision{Since LTS is based on
flattening \julienRevision{(and as such modifies data values)}, it
introduces visual flat-plateau
artifacts~(\autoref{fig_IllustrationOfSimplificationStrategies}).}
\jonasRevision{This can be problematic if one wants to extract
geometrical features within the simplified regions, such as the intra-cellular
features of \autoref{fig_cell2}. Such geometries
should be extracted based on the original data, where the simplified data can be used as a mask for segmentation.}
Yet, these plateaus are needed to guarantee the removal of undesired
\julienRevision{topological}
features in level-set based segmentations~(\autoref{fig_TDAforDifferentSimplificationStrategies}),
and LTS can be combined
with numerical techniques to provide smoother results if needed.
\julienRevision{Outside of the simplified areas,
the integral lines are left unchanged, since simplification is only applied
locally.
Thus,
the separatrices of the Morse-Smale
complex outside of simplified regions are also not impacted (in
\autoref{fig_cell2}, the separatrices of the simplified complexes, bottom, are
\jonasRevision{subsets} of separatrices of the unsimplified one, top left).
Within the simplified regions, however, the geometry of integral lines can
change.
Although the Morse-Smale complex will indeed only detect the authorized
critical points~\cite{ttk}, the simplification may consequently have an impact
on how separatrices connect them, if they traverse simplified regions, which
requires more detailed investigation in future work.}

LTS focuses on extremum-saddle pairs and does not
support saddle-saddle pair removal.
However, from our experience,
the extrema of a scalar field are the topological objects that users
investigate in priority for feature extraction.
\jonasRevision{Moreover, if extrema are not selected based on persistence, the value of the remaining critical points may change after simplification.}
This happens for instance if the only minimum
to preserve is initially located higher
than the only maximum to preserve, in which case the algorithm will change
their values to satisfy the input constraints.
\jonasRevision{Finally, in our experiments we observed that computing
abstractions on
\julienRevision{the} pre-simplified
\julienRevision{data}
seems to be equivalent to
post-simplifying the topological abstractions themselves---e.g., the merge trees
and Morse-Smale complexes of the simplified fields are identical to the pruned
abstractions of the original field---but this observation needs to be further
evaluated in future work.}

\section{Conclusion}
\label{sec_conclusions}

This paper described a combinatorial approach for the localized topological simplification (LTS) of scalar data.
Given a PL scalar field~$f$ and
a
selection of extrema to preserve, LTS
transforms~$f$ to a new PL scalar field~$g$ via localized iterative flattening,
such that~$g$ is close to~$f$ and only exhibits the selected set of extrema.
LTS significantly accelerates an essential part of topological
data analysis by reducing the time
spent on data pre-simplification by \julienRevision{up to} an order of 
magnitude \julienRevision{in our experiments} (\speedup{36}).
In many instances, this brings the execution time of TDA pipelines from minutes 
down to a few seconds, which enables the interactive exploration of 
simplification parameters.
Although LTS
scales well in the presented experiments,
the tested implementation is optimized for
workstations.
Next, the code needs to be optimized for larger, shared-memory, high-performance
machines.
\julienRevision{Extensions to GPU computation will also be
considered.}
\julienRevision{For} larger problems, one can investigate distributed 
parallelism, for which the localized nature of LTS is also expected to be 
beneficial.
Another line of research is the removal of saddle-saddle pairs, which has been 
reported to be NP-hard in general~\cite{AttaliBDGL13}, and 
\julienRevision{thus}
heuristic 
approaches need to be investigated.

\vspace*{0.5em}
\section*{Acknowledgments}
\small{This work was supported by the U.S. Department of Homeland
Security under Grant Award 2017-ST-061-QA0001 and 17STQAC00001-03-03, and the
National Science Foundation Program under Award No. 1350573.
The views and conclusions contained in this document are those of the authors
and should not be interpreted as necessarily representing the official policies,
either expressed or implied, of the U.S. Department of Homeland Security.
This work was also partially supported by the European Commission grant
ERC-2019-COG \emph{``TORI''} (ref. 863464), and the German research foundation
(DFG) through~the~IRTG~2057.
Julien Tierny would like to dedicate this paper to his son Marvin.
}

\newpage
\bibliographystyle{abbrv-doi}
\bibliography{main}

\end{document}